\documentclass{pinchcr}   

\usepackage{graphicx}        
\usepackage{amsmath}
\usepackage{bm}
\usepackage{epsfig,amsmath,amssymb,graphics,color,calc}

\title{Kinetically Constrained Models}

\author{Juan P. Garrahan}
\affiliation{School of Physics and Astronomy, University of
Nottingham, Nottingham NG7 2RD, U.K.}

\author{Peter Sollich}
\affiliation{King's College London, Department of Mathematics,
Strand, London WC2R 2LS, U.K.}

\author{Cristina Toninelli}
\affiliation{LPMA -- Univ.Paris VI-VII, CNRS UMR 7599, Case courrier
188, 4 Place Jussieu, 75252 Paris Cedex 05, France}

\begin{document}

\maketitle

\preface
In this chapter we summarize recent developments in the study of
kinetically constrained models (KCMs) as models for glass formers.
After recalling the definition of the KCMs which we cover we study
the possible occurrence of ergodicity breaking transitions and
discuss in some detail how, before any such transition occurs,
relaxation timescales depend on the relevant control parameter
(density or temperature). Then we turn to the main issue: the
prediction of KCMs for dynamical heterogeneities. We focus in
particular on multipoint correlation functions and susceptibilities,
and decoupling in the transport coefficients. Finally we discuss the
recent view of KCMs as being at first order coexistence between an
active and an inactive space-time phase.

\chapter{Kinetically Constrained Models}

\section{Motivation}

\index{kinetically constrained models}Kinetically constrained models
(KCMs) are simple lattice models\index{lattice models} of glasses.
They furnish a perspective on the glass transition problem which has
its origin in the work of Glarum \shortcite{GLARUM:1960ul},
Anderson\shortcite{Anderson-LH} and coworkers
\shortcite{PalmerSAA84}, and Andersen and coworkers \shortcite{FA}.
This perspective assumes that most of the interesting properties of
glass forming systems are dynamical in origin, while thermodynamics
plays a very limited role. KCMs tend to have simple and
uninteresting thermodynamics, typically that of a non-interacting
lattice gas.  In contrast, they display rich dynamical behaviour as
a consequence of kinetic constraints. This combination of simple
thermodynamics and locally constrained dynamics is often assumed to
be the result of coarse-graining of a dense molecular
system\shortcite{GarCha03}: dense fluids are structureless at
distances beyond the molecular length, but interatomic forces at
high densities are highly constraining, giving rise to local
restrictions in the dynamics.  As such KCMs are meant as models of
glass forming systems at high densities or low temperatures, and aim
to capture their dynamical behaviour for motion beyond the
inter-molecular distance and for long times.

KCMs generally use local constraints. Nevertheless, as we discuss
below, these give rise to collective dynamics due to a form of
dynamical frustration: at low temperatures/high densities there is a
conflict between the scarcity of excitations/vacancies and the need
for them to ``facilitate'' local motion, leading to hierarchical and
cooperative relaxation.
Note that this frustration, in contrast with e.g.\ the random
first-order approach \shortcite{Lubchenko:2007fk}, does not arise
from quenched disorder, an ingredient which at any rate is not
trivial to justify in models of real liquids.
A further distinction from the latter approach is that KCMs offer a
``non-topographic'' \shortcite{Berthier:2003lk} view of the glass
transition problem. By this we mean that it is not a change in the
topographic structure of the potential energy landscape (such as a
transition between a saddle-dominated and a minima-dominated regime)
that determines glassy relaxation, but a change in the degree of
connectivity of the configuration space. Indeed, due to the presence
of the constraints, the effective connectivity of this space depends
on temperature/density, as we discuss in more detail in section
\ref{breaking}. This picture provided by the KCM approach is
appealing as an effective description of the physics for example in
the case of hard spheres, where all allowed configurations have the
same energy. For a more detailed discussion on the non-topographic
KCM approach versus the energy landscape scenario (and additional
references on the latter) we refer the reader to
\shortcite{Berthier:2003lk,Whitelam:2004wt}.
Further information on disorder or landscape based approaches, along
with other explanations of glassy behavior that invoke a
thermodynamic transition, can be found in Chapter 1 of this book.

KCMs are simple enough to allow for detailed analysis.  The current
interest in KCMs originates in the fact that they exhibit explicit
mechanisms for super-Arrhenius slowdown and stretched relaxation
\shortcite{Sollich:1999tg} as a consequence of local, disorder free
interactions, and without the emergence of finite temperature
singularities \shortcite{ToninelliBF04}. At the same time they
provide a natural explanation \shortcite{Garrahan:2002qp} for the
phenomenon of dynamical heterogeneity (for reviews see
\shortcite{Ediger00,Glotzer00,Andersen:2005fr}).  KCMs are to the
constrained dynamics/facilitation\index{facilitation} perspective
\shortcite{PalmerSAA84,FA} what the random energy model and the
$p$-spin spin glass are to the random-first order transition
approach \shortcite{Lubchenko:2007fk}. Insights from the analysis of
KCMs allow one to construct a comprehensive theoretical picture of
the glass transition problem (see \shortcite{GC-ARPC} for a review)
which is quite distinct from other competing theories
\shortcite{GoetSjoe92,DebenedettiS01,Lubchenko:2007fk,Kivelson:2008lr,Cavagna:2009rt}.
Beyond their importance as models for describing glassy
phenomenology, KCMs are of interest for the mathematical community.
Indeed, even though they belong to the class of interacting particle
systems with Glauber and Kawasaki dynamics, the rate at which
elementary moves (birth/death or jump of particles) occur may
degenerate to zero due to the presence of the kinetic constraints.
This prevents the use of the standard probabilistic tools developed
for such systems. Furthermore it gives rise to peculiar phenomena
including the presence of several invariant measures, ergodicity
breaking transitions \shortcite{TBspiral1}, unusually long mixing
times \shortcite{CMRT} and aging phenomena \shortcite{FMRT}.

 The most recent comprehensive review of KCMs is that of Ritort and
Sollich \shortcite{RitSol03} which covers the literature until
roughly the end of 2001. Earlier surveys can be found in Refs.\
\shortcite{Jaeckle86,Fredrickson88,Palmer89}.  The aim of this
chapter is to briefly describe the developments in the last decade
or so.  It is organised as follows. In Section~\ref{models} we
introduce the KCMs that we will cover, grouping them according to
whether they are non-conservative (Glauber dynamics) or conservative
(Kawasaki dynamics). Section~\ref{breaking} is devoted to a
discussion of probably the most basic mathematical manifestation of
glassiness in KCMs, i.e.\ whether they possess \index{ergodicity
breaking transitions}ergodicity breaking transitions. Beyond such a
transition, the relaxation time to equilibrium is effectively
infinite. Physically, of course, this may not be distinguishable
from a finite but very long time, and so we consider in
Section~\ref{time} how relaxation timescales in KCMs depend on the
relevant control parameters (temperature/density). These times do
indeed generically increase extremely fast, in particular in the
so-called cooperative models. In Section~\ref{hetero}, finally, we
turn to the predictions of KCMs for \index{dynamical
heterogeneity}dynamical heterogeneity  in glasses. Here the models
provide a appealingly intuitive picture that is directly based on
considering the dynamics in real (rather than e.g.\ Fourier mode)
space. We cover definition and predictions for multi-point
correlation functions and susceptibilities, decoupling in the
temperature or density-dependence of transport coefficients, and
finally the view of KCMs as being at a first-order coexistence
between two different dynamical (space-time) phases. We conclude in
Section~\ref{conclusion} with a summary and a critical discussion of
the advantages and drawbacks of KCMs. We also provide a brief
comparison with alternative approaches, and finally an outlook
towards future work.

\section{The models}
\label{models}

The majority of KCMs are defined as stochastic lattice models with
binary degrees of freedom, and it is on such models that we focus in
this chapter.  We do not attempt to give here an exhaustive survey
of the wider range of models studied previously, but refer to this
for the review listed in the
introduction~\shortcite{RitSol03,Jaeckle86,Fredrickson88,Palmer89}.
For our present purposes, then, a KCM has on each lattice site $i$
an occupation variable $n_i\in\{0,1\}$, and the collection of the
$n_i$ defines the overall configuration $\bm{n}$. {\em Conservative}
KCMs are lattice gases where the $n_i$ indicate the presence
($n_i=1$) or absence ($n_i=0$) of particles. The dynamics follows a
continuous-time Markov process that consists of a sequence of
particle jumps, subject to kinetic constraints\index{kinetic
constraints} as explained below. The total particle number $\sum_i
n_i$ is conserved.

We will also discuss {\em non-conservative} KCMs. These are
motivated by a conceptual coarse-graining to a length scale several
times the particle diameter of the underlying physical glass that is
being modelled.  Each lattice site $i$ then represents a small
region of material containing at least a few particles, and one sets
$n_i=1$ (respectively $n_i=0$) if the density in this region is
above (respectively below) a certain threshold where local
re-arrangements -- of the particles inside the element -- become
possible. This representation no longer contains the precise values
of all local densities, and consequently in non-conservative KCMs
moves that change $\sum_i n_i$ are allowed.  The reader should be
aware that in much of the literature a reverse convention for the
two states of $n_i$ is used, with $n_i=1$ standing for high
mobility, i.e.\ low density, and vice versa for $n_i=0$.  (These
states are then often further identified as up- and down-spins, but
we will avoid this terminology.) We opt for the version discussed
above as it makes for a unified discussion of conservative and
non-conservative KCMs: $n_i=0$, an empty or low-density site, is
mobile in both contexts and facilitates the dynamics on neighbouring
sites. For brevity we use the terms ``empty'' and ``low-density''
interchangeably below, and similarly for ``occupied'' and
``high-density''. When we talk about the density of the system, we
correspondingly mean the fraction of sites with $n_i=1$, both in
conservative and non-conservative KCMs.

The key property of all KCMs is that in order to perform a move the
configuration must satisfy a local constraint which usually
corresponds to requiring a minimal number of empty sites in some
appropriate neighbourhood. This represents the physical intuition
that, when particles re-arrange in a glass, motion in any given
region requires the presence of mobile regions around
it~\shortcite{FA,KA}. When the density of these mobile or
facilitating sites decreases, the dynamics slows down. Another
property shared by all models is that the rates satisfy detailed
balance with respect to (w.r.t.) a Boltzmann distribution that
factorizes over sites (mathematically, a Bernoulli product measure),
so that there are no static interactions, beyond the effective hard
core repulsion implemented by the restriction $n_i\in\{0,1\}$. As
explained in the introduction, this idealization means that KCMs can
also be viewed as an attempt to find out how much of glass
phenomenology can be explained on purely dynamical grounds, without
recourse to e.g.\ static phase transitions.

KCMs as described above are plainly quite simplistic compared to
more realistic interacting atomic or molecular systems. As we shall
explain, however, for appropriate choices of the constraints they
display a behavior which is in agreement with the broad
phenomenology of glass forming liquids including super-Arrhenius
slowing down of the dynamics, stretched exponential relaxation,
dynamical heterogeneities, aging phenomena and ergodicity breaking
transitions.



\subsection{Facilitated spin models: FA, East and Spiral models}

We next describe the non-conservative KCMs that we will consider.
Because mobile ($n_i=0$) sites facilitate motion on neighbouring
sites, and because binary degrees of freedom with non-conservative
dynamics can be interpreted as spins, such models are also called
``facilitated spin models''\index{facilitated spin models}. In these
models, the rate for changing the state of site $i$ is
$f_i(\bm{n})[(1-\rho)n_i+\rho(1-n_i)]$, where $f_i$ depends on the
configuration $\bm{n}$ in a finite neighbourhood of $i$, but {\em
not} on $n_i$ itself. It is then immediate to verify that detailed
balance holds w.r.t.\ a Boltzmann distribution with energy function
$-\sum_i n_i$ and inverse temperature $\beta$ linked to the density
by $\rho=1/(1+e^{-\beta})$: in equilibrium we have $n_i=1$ with
probability $\rho$ and $n_i=0$ with probability $1-\rho$,
independently at each site.  In general $f_i$ is chosen to be
nonzero (allowed move) when an appropriate neighbourhood of $i$
contains a minimal number of empty sites, otherwise it is zero
(forbidden move). As the temperature $1/\beta$ decreases, the
density $\rho$ of occupied sites increases and the density
$q=1-\rho=1/(1+e^{\beta})$ of empty sites decreases: the dynamics
must then slow down.

Non-conservative KCMs can be divided into two classes:
non-cooperative and cooperative models. For the former it is
possible to construct an allowed path -- a sequence of
configurations linked by transitions with nonzero rates -- which
completely empties {\em any} configuration provided that it contains
somewhere an appropriate finite cluster of empty sites, the {\em
mobile defect}. For cooperative models, no such finite mobile
defects exist. As we will detail in Section~\ref{time},
non-cooperative and cooperative models display respectively an
Arrhenius and super-Arrhenius slowing down.

We now give the definitions of some specific models which cover
these two different classes of behaviour. The first class of
non-conservative KCMs was proposed by Fredrickson and Andersen (FA)
\shortcite{FA}, hence the name \index{Fredrickson-Andersen
models}Fredrickson-Andersen models.
 The simplest of these models, which we write as FA-1
(``one-spin facilitated FA''), allows a change of state at site $i$
only if at least one of the nearest neighbours is empty:
$f_i(\bm{n})=1$ if $\sum_{j\sim i} (1-n_j)>0$, $f_i(\bm{n})=0$
otherwise, where the sum runs over the nearest neighbours $j$ of
site $i$. It is easy to check that the presence of a single empty
site allows one to empty the whole lattice: the model is
non-cooperative. For theoretical calculations it is often easier to
define $f_i$ somewhat differently, as $f_i(\bm{n})=\sum_{j\sim i}
(1-n_j)$. This leaves the kinetic constraints enforced by vanishing
rates as they are and just changes the rates for allowed moves when
more than one empty neighbour is present; as such, it makes no
qualitative difference to the behaviour of the model.


One can similarly define $m$-constrained FA models, FA-$m$, by
setting $f_i(\bm{n})=1$ if at least $m$ neighbouring sites of $i$
are empty, and $f_i(\bm{n})=0$ otherwise. These models are normally
considered on hyper-cubic lattices of dimension $d$, with $2\leq
m\leq d$~\shortcite{FA}. As can be checked directly, it is not
possible in any of these models to devise a finite  cluster of empty
sites which always lets one empty the entire lattice. Consider for
example the case $d=2$, $m=2$ in infinite volume
 and focus on a configuration which
contains two adjacent infinite rows of occupied sites. Even if the
rest of the lattice is completely empty, none of the sites in these
two rows can ever change state because each has at most one empty
neighbour. Thus it is not possible to devise a mobile defect which
can unblock (empty) every configuration and the model is
cooperative. The same conclusion applies on a finite lattice, e.g.\
with periodic boundary conditions. The restriction on $m$ comes from
the fact that, if $m>d$, it is possible to construct {\em finite}
structures of occupied sites that are blocked, i.e.\ can never
change state.  They are therefore not suitable to describe the slow
dynamics close to a glass or jamming transition because a finite
fraction of the system is jammed at {\em any} density.

Another example of a cooperative model is the one-dimensional
\index{East model}East model \shortcite{JE}. In this case the
constraint requires for an allowed move that the left nearest
neighbour be empty, that is $f_i=1-n_{i-1}$. Note that on a finite
chain the presence of a single empty site at the left boundary lets
one empty the entire chain; but this is not a mobile defect because
it will work only when it is found in a specific position. In a
chain that is full except for a finite cluster of empty sites in a
generic location, the part of the chain to the left of the cluster
cannot be emptied, and so the model is cooperative.  One can show
that due to the directed nature of the constraint the relaxation
involves the cooperative rearrangements of increasingly large
regions as $q$ becomes small. This leads to super-Arrhenius
behaviour of relaxation time scales\shortcite{SE,SolEva03}. (The
East model shows up a slight shortcoming of our definition of
co-operativity: with periodic boundary conditions, we would formally
label this model non-cooperative because all sites can be emptied
starting from a single empty site. Physically, the model is clearly
co-operative independently of the boundary conditions.) We discuss
below also higher-dimensional generalizations of the East model,
e.g.\ in $d=3$ the North-East-Front (NEF) model, where a local
change of state is allowed if at least one of the neighbouring sites
to the North or East or front is empty.

\begin{figure}[htp]
\includegraphics[width=0.99\columnwidth]{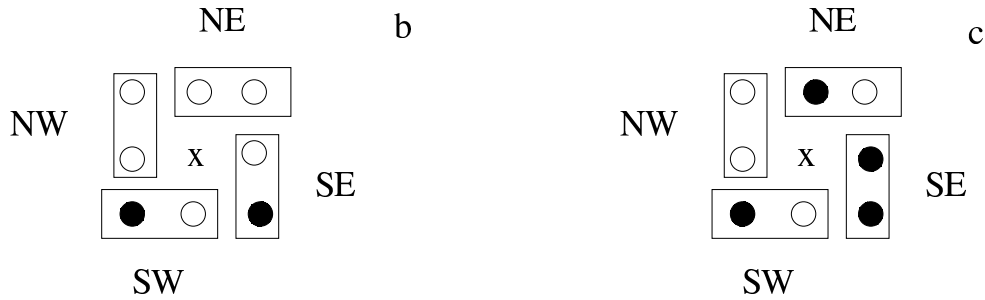}
\caption{Site $i$ and its NE, NW, SE and SW neighbours.  The
  constraint is (is not) verified at the central site in case (a) (in
  case (b)).
\label{BPvoids} }
\end{figure}

The models described above have been closely studied since they were
proposed in the 80s~\shortcite{FA} and 90s~\shortcite{JE}, with some
work continuing into the present decade~\shortcite{RitSol03},
particularly regarding the non-equilibrium behaviour (see
Section~\ref{sec:non-eq}) and detailed comparisons with glass
phenomenology~\shortcite{GarCha03,Jung:2004oj,BerChaGar05,ToninelliWBBB05,Chandler:2006gd,Elmatad:2009lr}
In parallel, progress has come from the development of new
KCMs~\shortcite{GarCha03}. We discuss below one of these, which
reproduces in particular the mixed character of the glass
transition, where diverging lengthscales characteristic of second
order transitions combine with a first order jump in the fraction of
frozen degrees of freedom.
This two-dimensional \index{spiral model}Spiral
model~\shortcite{TBFreply,TBspiral1,TBspiral2} is defined as
follows. Consider a square lattice and, for each site $i$, divide
its first and second neighbours into the North-East (NE), South-West
(SW), North-West (NW) and South-East (SE) pairs, as illustrated in
Fig.~\ref{BPvoids}.
Then the constraint for a move to occur at $i$ is that ``both its NE
{\em and/or} both its SW neighbours should be empty'' {\em and}
``both its SE {\em and/or} both its NW neighbours should be empty''
(see Fig.~\ref{BPvoids}); in other words, $f_i(\bm{n})=1$ if this
condition is satisfied and $=0$ otherwise. Stated in a more
geometrical form, the constraint requires that all four sites in one
of four contiguous sets of neighbours (NE and SE, or SE and SW, or
SW and NW, or NW and NE) have to be empty. The motivation for the
somewhat involved form of this constraint will become clear when we
analyse the dynamical transition in the model below.
We note that before the Spiral model a somewhat more involved choice
of the constraints (the so called Knights model) was proposed
\shortcite{knight, TBFreply} which leads to the same mixed
first/second order transition. Other choices in the same spirit have
also been investigated since, see for example \shortcite{Jeng}.

\subsection{Kinetically constrained lattice gases: KA and TLG models}

Conservative KCMs are defined in a similar spirit to their
non-conservative analogues, but with dynamics that conserves total
particle number. A particle at $i$ attempts at a fixed rate ($=1$
without loss of generality) to jump to any empty nearest neighbour
site $j$. The rate for this move is thus $n_i(1-n_j)f_{ij}(\bm{n})$.
The factor $f_{ij}(\bm{n})$ again implements a kinetic constraint
and is chosen not to depend on the configuration at sites $i$ and
$j$.  The resulting dynamics preserves the number of particles, and
so detailed balance on finite lattices is satisfied w.r.t.\ the
Boltzmann distribution which is uniform on configurations with the
appropriate fixed particle number. By linearly combining
distributions with different particle numbers one can of course also
create grand-canonical Boltzmann distributions where each site
independently contains a particle with some probability $\rho$.
Detailed balance is also satisfied w.r.t.\ these distributions, and
they extend naturally to the limit of an infinite lattice.
%
%

Again, we can classify conservative KCMs into non-cooperative and
cooperative models.  For the former it is possible to construct a
finite group of empty sites, the {\em macrovacancy}, such that for
{\em any} configuration the macrovacancy can be moved all over the
lattice and any jump of a particle to a neighbouring empty site can
be performed when the particle is adjacent to the macrovacancy.
Clearly the baseline lattice gas in which there are no kinetic
constraints, which in $d=1$ is the symmetric simple exclusion
process (SSEP), is non-cooperative and the minimal macrovacancies
are just isolated empty sites.

Chronologically the first conservative KCM is the
\index{Kob-Andersen model}Kob-Andersen model (KA) \shortcite{KA},
which has cooperative dynamics. Here a particle can jump to a
neighbouring site only if both in the initial and final position at
least $m$ of its nearest neighbouring sites are empty. The original
KA model had $m=3$, with particles on a cubic lattice, but one can
similarly define KA-$m$ models on hyper-cubic lattices of dimension
$d$ and for different values of the parameter $m$, with $2\leq m
\leq d$. The restrictions on $m$ arise from the fact that $m=1$
corresponds to an unconstrained lattice gas, while a model with
$m>d$ has a finite fraction of frozen particles at any density (as
for FA models in the same parameter regime). Similarly to the case
of FA-$m$ models with $m\geq 2$, one can directly check that all
KA-$m$ models are cooperative.
%
%
On other lattices, KA models can be non-cooperative. Consider for
example the KA model with $m=2$ on a triangular lattice. Here one
can check that a ``dimer'' of two neighbouring empty sites forms the
required macrovacancy, which can move across the lattice in a
tumbling motion even when all other sites are occupied.
Two other KCMs on triangular lattices were introduced in
\shortcite{JaecKro94} and have more recently been analysed in
\shortcite{HedGar07}. For the ($1$)-TLG a particle can move from
site $i$ to a neighbouring site $j$ if at least one of the two
mutual neighbours of $i$ and $j$ is empty. For the ($2$)-TLG a
particle can move from $i$ to $j$ if {\em both} the two mutual
neighbour sites are empty. It is easy to verify that the ($1$)-TLG
is non-cooperative, with -- as for KA with $m=2$ on the same lattice
-- a dimer of vacancies being a macrovacancy. The ($2$)-TLG, on the
other hand, is cooperative: a chain of particles occupying an entire
row of the lattice, for example, can never be destroyed, thus it is
not possible to construct a macrovacancy.


\section{Ergodicity breaking transitions}
\label{breaking}

As explained above, the equilibrium distribution in KCMs is trivial
because it factorizes over sites. In particular, then, no
equilibrium phase transition can occur. On the other hand, the
presence of constraints might induce transitions of purely dynamical
type: detailed balance alone does not guarantee that the
distribution over configurations will converge for large times to
the Boltzmann equilibrium distribution. To see this, return to the
example of the FA-2 model on a square lattice ($m=d=2$), either
finite with periodic boundary conditions or infinite, and a starting
configuration $\bm{n}(0)$ which has two adjacent rows of occupied
sites.  Then since these sites are forever blocked, for any site $i$
within the two rows, $\lim_{t\to\infty}\langle n_i\rangle_t=1\neq
\langle n_i\rangle_{\rm eq} =\rho$, where $\langle \cdot\rangle_t$,
is the average over the stochastic dynamics which starts from
$\bm{n}(0)$ and $\langle \cdot\rangle_{\rm eq}$ is the equilibrium
average.

Motivated by this, we can ask whether convergence to the equilibrium
distribution is recovered at least in the thermodynamic limit: if we
sample an initial configuration $\bm{n}(0)$ from the equilibrium
distribution, then is the large time limit of the average of any
function $g(\bm{n})$ equal (with probability one)
 to its equilibrium average, i.e.\ is $\lim_{t\to\infty}
\langle g\rangle_t = \langle g\rangle_{\rm eq}$?  If this is the
case we will say that the system is {\em ergodic}.
The models we consider are all ergodic at sufficiently high density
$q=1-\rho$ of facilitating (empty) sites. If they do become
non-ergodic as $q$ is reduced, we call $q_c$ the critical density of
empty sites at which this transition occurs. As will be discussed
below, the models defined in the previous sections are ergodic at
any positive $q=1-\rho$, so that $q_c=0$, with the exception of the
Spiral model and FA models on Bethe lattices which display
ergodicity breaking transitions at $0<q_c<1$.

Owing to the factorized form of the equilibrium distribution,
ergodicity corresponds to the fact that in the thermodynamic limit
the configuration space is covered by a single irreducible set,
i.e.\ a set of configurations which are connected to each other by
allowed paths; for a proof see \shortcite[Proposition 2.4]{CMRT} and
\shortcite[Section 2.3]{KACMRT}. This in turn is equivalent to the
requirement that for any site $i$ (in non-conservative KCMs) or pair
of nearest neighbour sites $i,j$ (in conservative KCMs) there is an
allowed path which transforms the configuration into one where the
constraint is satisfied at $i$ ($i,j$). In other words, the
probability that any site $i$ belongs to a cluster of forever
blocked sites must vanish in the thermodynamic limit.

For {\em non-cooperative} KCMs one sees easily that ergodicity holds
at any density $\rho<1$, i.e.\ at any $q>0$. In the non-conservative
case, the probability of finding at least one mobile defect in an
equilibrium configuration goes to one in the infinite volume limit,
and then starting from this defect one can empty all sites. Thus,
any configuration is connected to the ``all empty'' configuration
with probability one. The situation is similar for non-cooperative
but conservative KCMs.


The case of {\em cooperative} KCMs is more delicate and here an
ergodicity breaking transition can occur. The non-conservative case
is again simpler, so we begin with this. In order to determine the
probability that a site belongs to a blocked cluster, consider the
following deterministic procedure: iteratively empty all sites for
which the constraint is verified until we reach either the
completely empty configuration, or one in which there is a
``backbone'' of mutually blocked occupied sites. Then it is easy to
verify that the sites that are blocked forever under the stochastic
dynamics of the actual KCM are precisely those that are empty at the
end of this deterministic procedure.
In other words, the problem of the existence of an ergodicity
breaking transition for cooperative non-conservative KCMs can be
reformulated as a percolation transition for the final configuration
of the above deterministic dynamics. For FA-$m$ models this
deterministic dynamics coincides with the well known algorithm of
bootstrap percolation and the results in \shortcite{AL,Schonman}
establish that $q_c=0$ on hypercubic lattices in any dimension $d$
and for any facilitating parameter $1\leq m\leq d$ (while trivially
$q_c=1$ for $m>d$ since finite blocked structures can occur). This
work disproved a long-standing conjecture from the original FA
paper~\shortcite{FA}, namely that an ergodicity breaking transition
would occur at some $q_c>0$.  Such a transition does take place,
however, when one considers FA models on a Bethe lattice as will be
explained in Section \ref{bethe}. An example of a finite-dimensional
model displaying such a transition is provided by the Spiral model
(see Section \ref{spiral}).

For cooperative conservative KCMs, the proof of ergodicity is more
involved. For example, for the KA-2 model on a square lattice
($d=2$),
we have shown that the irreducible set of configurations that has
unit probability in the thermodynamic limit is the one containing
all configurations which can be connected by an allowed path to a
configuration which has a frame of empty sites on the last shell
before the boundary~\shortcite{KAnoi}. Establishing that this set
has unit probability in the thermodynamic limit is more complicated
than in the corresponding non-conservative KCMs, i.e.\ FA models.
This is because there is no deterministic bootstrap-like procedure
which allows one to establish whether or not a configuration does or
does not belong to the irreducible set. Nevertheless, a formal proof
can be constructed~\shortcite{KAnoi,KACMRT}, and demonstrates that
the ergodicity breaking transition originally conjectured by Kob and
Andersen~\shortcite{KA} does not exist. Analogous arguments can be
constructed for the other choices of $m$ and $d$ \shortcite{KAnoi}
to rule out the occurrence of a transition.

\subsection{FA models on Bethe lattices}
\label{bethe}

As a simple example of a model that does have an ergodicity breaking
transition we consider next the FA-$m$ model on a Bethe lattice,
i.e.\ a random regular graph of connectivity $k+1$. Exploiting the
local tree-like structure of such a graph, it is easy to write a
self-consistent equation for the probability $P$ that a given site
is occupied and blocked, conditional on the fact that its ancestor
is occupied and blocked:
\begin{equation}\label{f-coreeq}
P = (1-q) \sum_{i=0}^{m-1} {k \choose i} P^{k-i} (1-P)^{i} .
\end{equation}
The factor $1-q$ is the probability for the given site to be
occupied; the sum gives the probability that at most $m-1$ of the
descendants of the site are empty so that it is in fact blocked. The
leading term on the r.h.s.\ for small $P$ is $O(P^{k-m+1})$. For
$m=k$, the largest value that does not produce finite blocked
structures, this is linear and so one gets a continuous transition
at $(1-q_c)m=1$. For all smaller $m$, the small $P$-increase is with
a higher power of $P$ and so a discontinuous transition results,
with $P=0$ for $q>q_{c}$ and $P=P_{c}+O (\sqrt{q_c-q})$ for $q<q_c$,
where $P_{c}>0$.  The mechanism behind this combination of a
discontinuous onset and a critical singularity has been analysed in
detail \shortcite{Mendes,Schwarz}. In particular, the singular
square root behavior is due to the extreme fragility of the infinite
spanning frozen or ``jammed'' cluster at the transition, and to the
existence of a related length scale which diverges as a power law.

\subsection{Spiral model}
\label{spiral}

\begin{figure}[htp]
\centerline{\includegraphics*[width=0.6\columnwidth,clip]{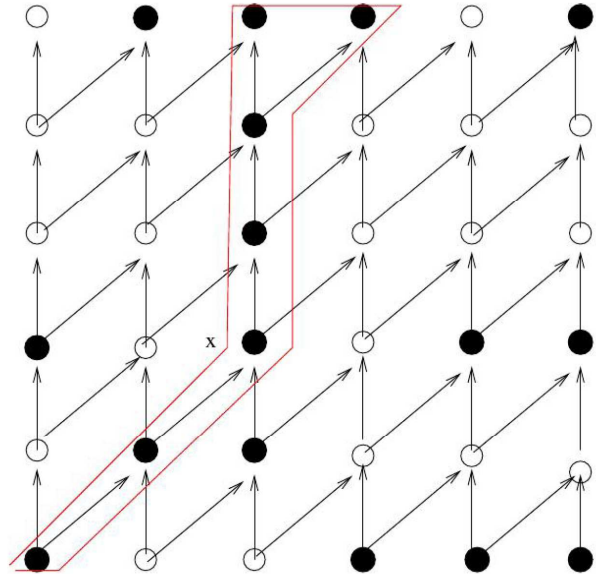}}
\caption{The tilted and squeezed two dimensional oriented lattice
  obtained by drawing arrows from each site to its
  neighbours to the North and East. The site marked by the cross belongs to an oriented occupied cluster and is therefore blocked.
\label{tilted} }
\end{figure}

In this section we will explain the mechanism behind the ergodicity
breaking transition of the Spiral model \shortcite{TBFreply,
  TBspiral1,TBspiral2}.  Consider the directed lattice that is
obtained from the square lattice by putting two arrows from each
site towards the neighbours in the NE pair, i.e.\ pointing North and
North-East. The resulting lattice is a tilted and squeezed version
of a two-dimensional oriented square lattice (see Figure
\ref{tilted}). Therefore, if the density is larger than the critical
density of oriented site percolation (directed percolation, DP),
$\rho_c^{\rm DP}$, there exists a cluster of occupied sites which
spans the lattice following the direction of the arrows. Consider
now a site in the interior of this directed cluster (see Figure
\ref{tilted}): by definition there is at least one occupied site in
both its NE and SW neighbouring pairs, therefore the site is blocked
with respect to the constraints of the Spiral model. Thus for
$\rho>\rho_c^{\rm DP}$ the system contains one or more blocked
clusters and is therefore non-ergodic. This suggests that the
ergodicity breaking transition occurs at $q_c = 1-\rho_c^{\rm DP}$,
but the argument so far does not exclude a transition earlier, i.e.\
at a larger density $q$ of empty sites.  Indeed, since blocking can
occur along either the NE-SW or the NW-SE direction (or both), the
presence of a blocked cluster does not imply that there is a
directed path through the lattice as in the DP argument. To
establish that, nevertheless, blocked clusters do not occur for
$q>1-\rho_c^{\rm DP}$, one shows that empty regions of linear size
much larger than the parallel length of DP, $\xi_{\parallel}$, act
as ``critical defects'': starting from any such empty region we can
very likely empty the whole lattice.
Because $\xi_{\parallel}$ is finite, so is the size of these
defects.  In the thermodynamic limit at least one such defect will
occur in the system, and so for $q>1-\rho_c^{\rm DP}$ any
configuration can be completely emptied: the system is ergodic. This
confirms that indeed $q_c = 1-\rho_c^{\rm DP}$.
%


An important feature of the ergodicity breaking transition in the
Spiral model is that it is discontinuous, in the sense that the
fraction of blocked sites jumps to a nonzero value at $q=q_c$. This
is relevant for the connection to real glasses, because it implies
that two-point correlation and persistence functions must display a
plateau in their time-dependence when the transition is approached
($q>q_c$), as is observed experimentally. Earlier models with a
transition at a finite $q_c$, e.g.\ a two-dimensional generalization
of the East model~\shortcite{ReiMauJaec92}, develop only a fractal
cluster of blocked particles at the transition, which occupies a
vanishing fraction of the system. The fact that in the Spiral model
blocked clusters are compact at $q_c$ follows by a direct
construction of blocked structures
\shortcite{TBFreply,TBspiral1,TBspiral2} and is a consequence of the
presence of the two transverse blocking directions in the
constraints, and of the anisotropy of DP.

\subsection{Summary: Presence/absence of ergodicity breaking transition}
\label{classi}


For the sake of clarity let us summarize the results of this section
by listing the KCMs we have discussed according to whether or not
they display an ergodicity braking transition at a non-trivial
(different from 0 or 1) critical defect density $q_c$.

The models that do display such a transition are the FA and KA
models on Bethe lattices, the Spiral model and the North-East model.
The latter was not mentioned above because it is not of direct
interest for modelling the glass transition, due to the continuous
character of its ergodicity breaking transition. It is a spin
facilitated model on a square lattice with the constraint requiring
both the North and East neighbour to be empty. It is easy to verify
along the lines of the argument in the previous subsection that the
North-East model as defined by this constraint displays a transition
at the critical density of oriented percolation.

The models that do not display a transition are: the FA-$m$ and
KA-$m$ models for any choice of $m$ and on hypercubic lattices in
any finite dimension $d$; the East model; and the $(1)$-TLG and
$(2)$-TLG models.

\section{Bulk dynamics of KCMs}
\label{time}

\subsection{Glassy time scale divergences and static length scales}
\label{subsec:time}

For KCMs to be useful as models of physical glasses they need to
satisfy the basic requirement of dynamical slowing down as the
glassy regime approached. In our case this corresponds to increasing
density $\rho$, or decreasing density of empty sites $q$. We will
therefore now give an overview of numerical and analytical results
on how relaxation time scales $\tau$ diverge in KCMs when $q$
approaches $q_c$. In concrete terms, we will take $\tau$ as the
typical time in the relaxation of density-density correlation and
persistence functions. In cooperative KCMs, $\tau$ turns out to be
connected to statically defined blocking lengths, as we also
discuss.

Starting with FA-1 models, relaxation occurs via the effective
diffusion of empty sites: an empty site facilitates the emptying of
a neighbour site, with rate $q=1-\rho$, and the original site can
then fill and does so with probability $1/2$ before the new empty
site does. Empty sites can thus diffuse freely, with an effective
diffusion constant $q/2$. This suggests that $\tau$ should grow as
an inverse power of $q$. Because $q\simeq \exp(-\beta)$ at low
temperatures $1/\beta$, this corresponds to Arrhenius scaling.
Indeed, by an exact mapping to a diffusion-limited aggregation model
one can derive~\shortcite{Jack-Mayer-Sollich} that $\tau \sim 1/q^z$
with $z=3$ in $d=1$ and $z=2$ in $d\geq 2$. The $d=1$ result is
simple to understand: empty sites are typically a distance $1/q$
apart, and relaxation requires diffusion across this distance with
the effective diffusion constant $q/2$, so $\tau\sim
(1/q)^2/(q/2)\sim 1/q^3$.

For FA-$m$ with $m>1$, an isolated empty site is unable to move on
its own and has to wait for a mobile defect, i.e.\ an appropriate
region of empty sites, to move cooperatively into its neighbourhood
and so facilitate its motion. As we will detail below, the typical
number of moves involved in this cooperative process increases as
$q\to 0$, thus a super-Arrhenius scaling of the relaxation time has
to result.  This has been confirmed by several numerical
investigations which have proposed different forms for the density
dependence of $\tau$ \shortcite{BH,FB,GPG}. In order to better
understand the cooperative mechanism let us consider for example the
case $m=d=2$.  An $\ell\times\ell$ square of empty sites can be
expanded by one lattice spacing in both directions, provided that at
least one empty site is present on two adjacent sides of the square.
This is very likely to be the case when $\ell\gg 1/q$.  On the other
hand, the probability of being able to empty all sites in a region
of size $1/q$ without using external empty sites can be shown to be
proportional to $\exp(-c/q)$ with $c$ a constant of order
one~\shortcite{AL,Ho}. This result is obtained by iterating the
process outlined above, multiplying the relevant probabilities at
each step: one can iteratively remove all the particles in any
region starting from its interior provided there is a central
$2\times 2$ square of empty sites and one additional empty site
somewhere on each of the four sides of each subsequent shell.
 In \shortcite{Reiter} it
was conjectured that relaxation occurs via the diffusion of these
critical defects leading to a relaxation time diverging as
$\tau_D/\rho_D$, with $\rho_D\simeq\exp(-c/q)$ the defect density
and $\tau_D$ their diffusion time.
The latter, as detailed in \shortcite[Section 6]{KAnoi}, should be a
sub-leading correction, giving $\tau\sim 1/\rho_D$ to leading order.
In principle one might think that this is not the optimal relaxation
mechanism and that other relaxation processes could be much faster,
avoiding the super-Arrhenius scaling. However, the results in
\shortcite{AL,Ho} also give the typical size of the incipient
blocked cluster to which any fixed site belongs and that has to be
eroded via successive moves from its boundary before the site in
qustion can be unblocked.
 This size, $L_c$, diverges as
$\exp(c/2q)$ thus providing (due to finite speed of propagation) a
lower bound on time scales of the form $\tau\geq \exp(c/2q)$ which
confirms the super-Arrhenius scaling. Note that $L_c^2$ scales as
the inverse of $\rho_D$, which (as the probability that a region of
site $1/q\times 1/q$ is a critical defect) is essentially the number
density of critical defects. Indeed, clusters of linear size larger
than $L_c$ are typically unblocked because they contain at least one
critical defect from which relaxation can occur. In
\shortcite{Schonman} the typical size of such incipient blocked
clusters was derived for generic $m$ and $d$ leading to $\tau$
growing at least as fast as $\exp^{\circ (m-1)}(c/q^{1/(d-m+1)})$.
Here $\exp^{\circ s}$ is the exponential iterated $s$ times, so that
the divergence of the time scale is extremely rapid for any $m\geq
3$.

For KA models the basic relaxation mechanism is very similar
\shortcite{KAnoi}, the main difference being that the critical
defects
are now regions in which empty sites and particles are arranged in
such a way that one can find an allowed path -- a sequence of
allowed moves -- to perform any nearest neighbour exchange.  As
shown in \shortcite{KAnoi,KAletter}, the properties as a function of
$q$ of these regions coincide with those for the FA models, leading
to the same estimates for the scaling of the relaxation time.

Relaxation processes in the Spiral model proceed in a broadly
analogous manner. But the size of the critical defects
that can expand further diverges already at $q_c$, proportionally to
the parallel length $\xi_\parallel$ of DP clusters. The typical size
of incipient blocked clusters grows much more quickly, as $L_c\simeq
\exp[c/(q-q_c)^{\mu}]$ with $\mu=\nu_{\parallel}(1-z)$, where
$\nu\simeq 1.73$ is the critical exponent of $\xi_{\parallel}$ and
$z\simeq 0.64$ the exponent relating the parallel and transverse
lengths of DP. As for FA models, since these clusters can be
unblocked only from the boundary, $\tau$ should diverge at least as
$L_c$.

Interestingly, in the other model we have considered that has an
ergodicity breaking transition at nonzero $q_c$, the FA model on a
Bethe lattice, simulation measurements of persistence and
correlation functions~\shortcite{Sellitto} show that relaxation
times grow only as power laws on approaching the transition, $\tau
\propto (q-q_{c})^{-\gamma }$ with $\gamma \simeq 2.9$.
Qualitatively, thus, the transition for FA models on Bethe lattices
has the characteristics of a mode-coupling theory
(MCT)~\shortcite{GoetSjoe92}
arrest transition.

Finally we discuss the East model, where the origin of the
super-Arrhenius time scale is possibly easiest to see.  The basic
relaxation mechanism can be understood in terms of the dynamics of
domains of occupied sites separated by empty sites. In the limit of
small $q$ one can then argue~\shortcite{SE} that the typical
relaxation time should scale as the minimal time required for an
empty site to facilitate the motion of the first empty site to its
right, which is typically at distance $1/q$. The optimal path to
create an empty site at a distance $d$ involves an energy barrier of
order $\log_2(d)$ (the minimum over all paths of the maximal number
of empty sites which we encounter along the path). Setting $d=1/q$
thus leads to the relaxation time estimate $\tau\sim
q^{-\log_2(1/q)}\sim\exp[\beta^2/\ln 2]$ \shortcite{SE}.  The above
argument does not take into account the behavior on scales smaller
than the typical distance between empty sites. Incorporating this
turns out to halve the coefficient in the exponent, to
$\tau\sim\exp[\beta^2/(2\ln 2)]$ \shortcite{CMRTjstat,CMRT}.

We have discussed so far only the overall timescale for the decay of
correlation or persistence functions in KCMs, but not the time
dependence of this decay. For the cooperative models one generically
finds stretched exponential forms, as seen experimentally, while
non-cooperative models can exhibit power law tails reflecting the
diffusive motion of the mobile defects. We refer
to~\shortcite{RitSol03} for an overview and quote only the example
of the East model, where one can show that the stretching becomes
extremely strong at low temperatures, with correlations and
persistence both decaying as scaling functions of
$(t/\tau)^{1/[\beta\ln 2]}$~\shortcite{SE,BuhGar01,SolEva03}.


\subsection{Some rigorous results}

In the recent years KCMs have been also analysed in the mathematical
community. We summarize here those rigorous results which help to
further understand the slow relaxation of KCMs, and have in some
cases corrected conjectures resulting from numerical simulations or
intuitive arguments.


The analysis of the large time behavior in the ergodic regime was
started by \shortcite{AD} where for the East model it was
established that $[1/\ln
2-o(1)]\ln^2(1/q)\leq\ln(1/\mbox{gap}(q))\leq[1/(2\ln 2)
+o(1)]\ln^2(1/q)$ with $1/{\mbox{gap}(q)}$ the inverse of the
\index{spectral gap}spectral gap of the Liouvillian operator
generating the dynamics. The latter, which in a finite system is
just the inverse of the smallest nonzero eigenvalue of the
transition matrix, represents the longest relaxation time for all
one-time quantities and so can be identified with the relaxation
time scale $\tau$ discussed above. The bounds of~\shortcite{AD}
quoted above then say that $\ln\tau$ scales for small $q$ as
$\ln^2(1/q)=\beta^2$, with a prefactor between $1/(2\ln2)$ and
$1/\ln2$, the upper bound being the naive estimate
of~\shortcite{SE}. These bounds were sharpened in~\shortcite{CMRT},
establishing that it is in fact the lower bound that gives the
correct asymptotics: written in terms of the gap, $\lim_{q\to
0}\ln(1/\mbox{gap})/\ln^2(1/q)=1/(2\ln 2)$.

Positivity of the spectral gap guarantees in particular exponential
convergence in the large time limit: for any $g(\bm{n})$ one has
$\langle g(\bm{n}(t))g(\bm{n}(0))\rangle - \langle
g(\bm{n}(0))\rangle^2$ $\leq$ $\mbox{const}\times
\exp(-2\,{\mbox{gap}}\,t)$, where $\langle \ldots \rangle$ is the
mean over the initial Boltzmann equilibrium at empty site density
$q$ and over the stochastic process governing the evolution in time.
 In \shortcite{CMRT} it was shown that positivity of the
spectral gap also guarantees exponential convergence of the
persistence function $P(t)$, which is the probability that a site
does not change its state during a time interval of length $t$:
$P(t)\leq\exp(-q\,\mbox{gap}\,t)+\exp(-(1-q)\mbox{gap}\,t)$.  In the
same paper, a multi-scale approach was developed which allows one to
prove positivity of the spectral gap in the whole ergodic region
$q>q_c$ for all the choices of constraints described in Section
\ref{models}. With this technique one can also
derive~\shortcite{CMRT} the following (sometimes optimal) bounds
when $q\downarrow q_c$. For FA-1 in $d=1$, $\tau\propto 1/q^3$; in
$d=2$, $1/q^2<\tau\leq \ln(1/q)/q^2$; and in $d\geq 3$,
$1/q^{1+2/d}<\tau\leq 1/q^2$. These bounds are in agreement with the
analytical results in \shortcite{Jack-Mayer-Sollich}, and confirm
that the initial findings \shortcite{WBG} in $d=2$ and $d=3$, based
on a mapping to DP, were incorrect. For the cooperative KCMs FA-2
and FA-3, results in \shortcite{CMRT} show that $\exp(q^{-1})\leq
\tau \leq \exp(q^{-2})$ and $\exp[\exp(q^{-1})]\leq\tau\leq
\exp[\exp(q^{-2})]$, respectively, thus establishing a
super-Arrhenius scaling compatible with \shortcite{BH,Reiter}.

For conservative KCMs with non-cooperative behaviour, the diffusive
scaling $1/\mbox{gap}=O(L^2)$ in a volume of linear size $L$ and the
positivity of the self-diffusion coefficient at any density was
established in \shortcite{Bertini}.  Moreover the hydrodynamic limit
was studied there for a special class of models leading to a porous
medium equation, namely a degenerate partial differential equation
$\partial_t\rho=\nabla(D(\rho)\nabla\rho)$ with a diffusion
coefficient $D(\rho)$ vanishing as a power law of $1-\rho$ when
$\rho\to 1$. The methods used to establish these results use as a
key ingredient the existence of a path between configurations which
allows any two particles to be exchanged, by exploiting the presence
of mobile macrovancancies. This approach, then, cannot be extended
to cooperative models. However, more recently a different technique
has been devised~\shortcite{KACMRT} that proves also for cooperative
models the diffusive scaling $1/\mbox{gap}=O(L^2)$ and establishes
in $d=2$ the diffusive decay in time $\sim 1/t$ of the
density-density autocorrelation function.  The self-diffusion
coefficient for a specific cooperative model, namely the KA-$m$, was
analysed in \shortcite{TB} where its positivity at any $q$ was
proved, modulo a conjecture on the behavior of a random walk in a
random environment.

\subsection{Non-equilibrium behaviour}
\label{sec:non-eq}

Due to space constraints we focus in this paper almost exclusively
on the equilibrium dynamics of KCMs. Non-equilibrium behaviour
results if, for example, a non-conservative KCM is prepared in a
configuration with a high density $q$ of empty sites, and then $q$
is reduced quickly. This corresponds to a sudden lowering of the
temperature $1/\beta$ and so mimicks quench experiments in real
glasses. If the final $q$ is low enough, i.e.\ sufficiently close to
$q_c$, the time scale $\tau$ for relaxation to the new equilibrium
will be very long, and aging will occur: the properties of the
system depend on the time elapsed since the quench. Aging can be
monitored via two-time response and correlation functions, and one
ask for example whether these are linked by a non-equilibrium
fluctuation-dissipation theorem with some effective
temperature~\shortcite{CriRit03}. Surprisingly, in some simple KCMs
such as the East and FA-1 models, aging at low $q$ is in fact easier
to analyse than the dynamics at equilibrium~\shortcite{SE,MaySol07}.
Work in this area up to late 2001 is summarized
in~\shortcite{RitSol03}, with more recent studies revealing a very
rich phenomenology in the aging of
KCMs~\shortcite{MayLeoBerGarSol06,LeoMaySolBerGar07,CorCug09}. There
has also been progress in rigorous approaches, with e.g.\
Refs.~\shortcite{FMRT,CMST} establishing the non-equilibrium
behavior of the East model derived in \shortcite{SE}.

\section{Dynamical heterogeneity and its consequences}
\label{hetero}

The main success of KCMs has been the ability to account, at least
qualitatively and sometimes also quantitatively, for many aspects of
dynamical heterogeneity (DH)
\shortcite{Ediger00,Glotzer00,Andersen:2005fr} observed in glass
forming systems.  DH emerges naturally in KCMs at high density or
low temperature \shortcite{GarCha03,Pan:2005lr}, and this is the
topic of this section.  We focus on the simpler KCMs, such as the
FA-1, East and TLG constrained lattice gases.
For these models results are easier to obtain both analytically and
computationally than for, say, the Spiral model or the FA-$m$ model
with $m>1$; nevertheless they display all the important physics. In
particular, the dynamical scaling properties of the East model are
close to those observed in realistic glass formers (see
e.g.~\shortcite{GC-ARPC}).

The simplicity of KCMs allows one to make detailed predictions about
the growth of dynamical correlation lengths and their scaling
relations to growing relaxation times
\shortcite{Garrahan:2002qp,ToninelliBF04,Whitelam:2004wo,Pan:2005lr,ToninelliWBBB05,Berthier:2005hb,Jack-Mayer-Sollich}.
These dynamical lengthscales are an indication of spatial
correlations that build up over time, so they are most easily
extracted from multi-point correlation functions, which we discuss
in Subsection \ref{multi}.  DH also tells us that the dynamics of
glass formers is fluctuation dominated.  A central consequence is
transport decoupling, the breakdown in standard transport relations
of liquid state theory that are obtained under the assumption of
homogeneous dynamics, such as that of Stokes-Einstein relating
self-diffusion rate to viscosity \shortcite{SwallenBME03}.  KCMs
provide a direct explanation for decoupling based on the response of
a moving molecule to the distribution of local relaxation timescales
in the host fluid \shortcite{Jung:2004oj}. Such motion can be
approximately quantified within a continuous-time random walk (CTRW)
formalism\shortcite{Berthier:2005hb}.  Decoupling and CTRW are
discussed in Subsection \ref{decoupling-section}.  Finally, the
observed DH is only a mesoscopic phenomenon: dynamical lengthscales
are always finite (and transient) at non-zero temperature or at less
than maximal density, that is, for $q>q_c$ (note that all models
discussed in this section have $q_c=0$).  Nevertheless, it is a
precursor to a fully-fledged non-equilibrium, or ``space-time'',
phase transition
\shortcite{Merolle:2005ix,Jack:2006ph,Garrahan:2007la}, which we
discuss briefly in Subsection \ref{spacetime-section}.

\subsection{Multipoint-correlations and susceptibilities}
\label{multi}

\begin{figure}[htp]
\includegraphics[width=0.9\columnwidth]{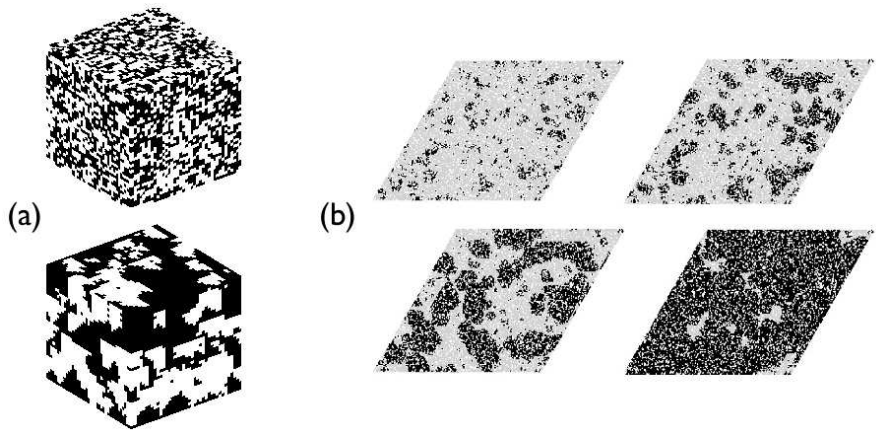}
\caption{Dynamical Heterogeneity in KCMs.  (a) Spatial distribution
of the local persistence in the NEF model, at time $t_{1/2}$ such
that $P(t_{1/2})=1/2$ (i.e., 50 of sites, shown in black, have
flipped by time $t_{1/2}$) at temperatures $T=1.0$ (top) and
$T=0.15$ (bottom) for a system of size $N=40^3$.  Adapted from Ref.~\protect\shortcite{Berthier:2005hb}, with permission.  (b) Same, but for
(2)-TLG model, at fixed density $\rho=0.77$, and varying observation
times, $t = 10^3,~10^4,~10^5,~10^6$.  From Ref.~\protect\shortcite{Pan:2005lr}, with permission.} \label{NEF-TLG}
\end{figure}

Figure \ref{NEF-TLG} illustrates DH in KCMs.  A convenient
observable to quantify local relaxation is the persistence field,
$p_i(t)=0,1$, where 0 indicates that site $i$ has changed its state
at least once up to time $t$, and 1 otherwise.  The ensemble
average, $P(t) \equiv \langle p_i(t) \rangle = \lim_{N \to \infty}
N^{-1} \sum_i p_i(t)$, is the persistence function discussed above.
Fig.\ \ref{NEF-TLG}(a) shows the persistence field $p_i(t_{1/2})$ in
a three-dimensional version of the East model, or NEF (for
North-East-Front model) \shortcite{Berthier:2005hb}, where $t_{1/2}$
is the time at which $P(t_{1/2})=1/2$, i.e., half the system has
relaxed and half has not. We see from Fig.\ \ref{NEF-TLG}(a) that
relaxation is heterogeneous: there is a clear spatial segregation of
sites which have relaxed, $p_i=0$ (colored black), from those that
have not, $p_i=0$ (colored white).  Relaxation dynamics is spatially
correlated.  Fig.\ \ref{NEF-TLG}(a) shows the persistence field at
two different values of $q$ (two different temperatures).  The
spatial extension of dynamic correlations increases with decreasing
temperature, and therefore with increasing relaxation time.  Note
that these dynamical correlations are unrelated to thermodynamic
correlations since the equilibrium measure of the NEF is trivial at
all temperatures. Fig.\ \ref{NEF-TLG}(a) shows similar persistence
field plots for a density conserving model, the two-vacancy assisted
(2)-TLG, but at a fixed high density $\rho$ and for different
observation times $t$.  This figure shows that DH is a transient
effect, with dynamical correlations becoming maximal at some
intermediate time $t^*$ (see below).

\begin{figure}[htp]
\includegraphics[width=0.9\columnwidth]{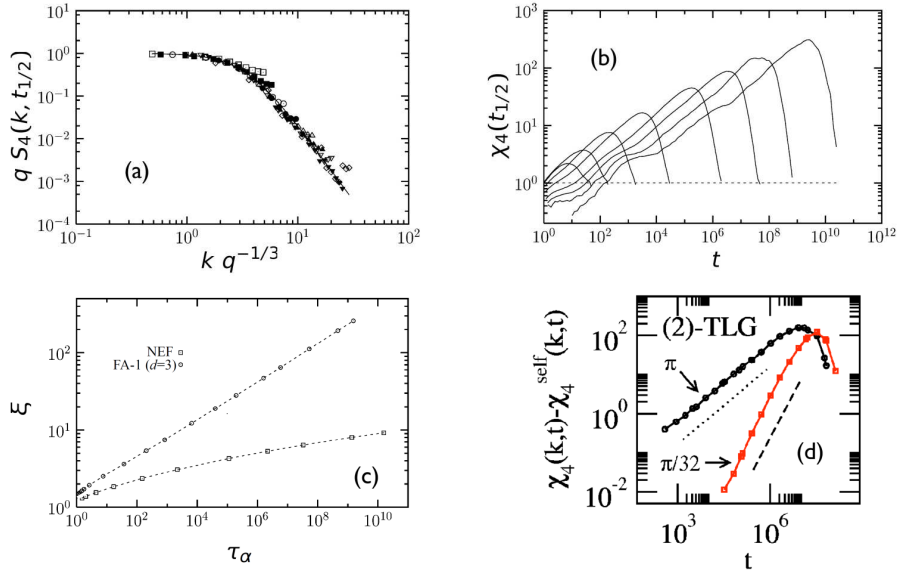}
\caption{Dynamical correlations and scaling.  (a) DH structure
factor, $S_4(k,t_{1/2})$ in the NEF model at various temperatures.
The data collapses under the scaling $S_4 \to q S_4$ and $k \to k
\xi$ with $\xi \sim q^{-1/3}$, suggesting the values $\gamma=1$ and
$\nu=1/3$ for the scaling exponents (see text). (b) Time dependence
of four point susceptibility $\chi_4(t)$ in the NEF.  Dynamical
fluctuations probed by this function are maximal at time $t^*$.  The
peak susceptibility $\chi_4^*=\chi_4(t^*)$ grows with decreasing
temperature.  The peak time is approximately the relaxation time of
the system, $t^* \approx \tau$, as extracted from the persistence
function.  (c) Scaling of DH correlation lengths to relaxation
times: FA-1 models follow a simple scaling law~\protect\shortcite{Jack-Mayer-Sollich} ($\tau \sim \xi^4$ for the
three-dimensional FA-1 shown in the figure), while for East-like
models $\tau \sim \xi^{z(q)}$, where the dynamical exponent $z(q)$
increases with decreasing $q$ (i.e.\ decreasing temperature)~\protect\shortcite{Sollich:1999tg}.  Adapted from
Ref.~\protect\shortcite{Berthier:2005hb}, with permission. (d) The behaviour
of four-point susceptibilities depend on the wavelength $k$ of the
observable which is used to define them. $\chi_4(k,t)$ is plotted
with its self-part, i.e.\ the terms with $i=j$ in its definition,
subtracted. From Ref.~\protect\shortcite{Chandler:2006gd}, with permission.
} \label{chi4}
\end{figure}

The extent of the dynamical correlations evident from Figure
\ref{NEF-TLG} can be quantified by means of multipoint functions
\shortcite{Garrahan:2002qp,Lacevic:2003dz,ToninelliWBBB05,Chandler:2006gd}.
Consider for example the ``four-point'' structure factor,
\begin{equation}
S_4(\vec{k},t) \equiv N^{-1} {\mathcal N}(t) \sum_{i,j} \left[ \langle
  p_i(t) p_j(t) \rangle - P^2(t) \right] e^{i \vec{k} \cdot (\vec{r}_i
  - \vec{r}_j)}
\end{equation}
This is the Fourier transform of the spatial correlation function of
the persistence field.  The factor ${\mathcal N}(t)$ is a convenient
normalization factor.  We adopt the choice ${\mathcal N}(t) = \left[
P(t) - P^2(t) \right]^{-1}$, which make $S_4$ equal to unity if all
the nonzero contributions come from ``self'' terms $i=j$.  Fig.\
\ref{chi4}(a) shows $S_4$ for the NEF.  This is the structure factor
of the DH pictures of Fig.\ \ref{NEF-TLG}.  It has the
characteristic shape of that of a system correlated over finite
distances.  The zero wave vector limit of $S_4$ gives the
``four-point'' susceptibility, $\chi_4(t)=S_4(k \to 0,t)$, which
provides an estimate of the correlation volume of DH.  Fig.\
\ref{chi4}(b) shows $\chi_4(t)$ for the NEF model as a function of
observation time $t$, at different temperatures.  Several things are
apparent. $\chi_4$ is non-monotonic in time, indicating that DH is a
transient phenomenon. It peaks at around the relaxation time of the
persistence function.  The peak value increases with decreasing
temperature: dynamical correlations increase with increasing
relaxation time.

Multipoint functions reveal the scaling properties of DH.  The
dynamic susceptibilities defined above have their peak, $\chi_4^* =
\chi_4(t^*)$, at times $t^*$ close to the relaxation times $\tau$ of
the corresponding persistence functions.  Scaling is controlled by
the distance to the dynamical critical point at zero concentration
of excitations or vacancies, so we expect $\chi_4^* \sim
q^{-\gamma}$ \shortcite{ToninelliWBBB05,Chandler:2006gd}.
Furthermore, we expect the four-point structure factor to behave as
$S_4 \approx \chi_4^* f(k \xi)$, where $f$ is a scaling function,
and $\xi \sim q^{-\nu}$ a correlation length for DH at times
$t=t^*$. These exponents -- and indeed scaling forms -- may vary
from model to model. For the simpler KCMs they can be calculated
analytically. Figure \ref{chi4}(a) shows the numerical collapse of
$S_4$ in the NEF for $\gamma = 1$ and $\nu = 1/3$
\shortcite{Berthier:2005hb}.  For other models, such as the FA-1
these exponents can be calculated analytically
\shortcite{ToninelliWBBB05}. For example, in $d=3$ one finds $\tau
\sim \xi^4$. This is most easily understood from the fact that the
upper critical dimension of the model is $d_c=2$, so exponents are
$d$-independent above this~\shortcite{Jack-Mayer-Sollich}. But in
$d=2$, the characteristic lengthscale is the distance between
vacancies of density $q$, $\xi \sim q^{-1/2}$, and the timescale is
related to this by the vacancy diffusion constant $q/2$ (see
Section~\ref{subsec:time}), giving $\tau \sim \xi^2/(q/2) \sim
\xi^4$.
%

\begin{figure}[htp]
\includegraphics[width=0.95\columnwidth]{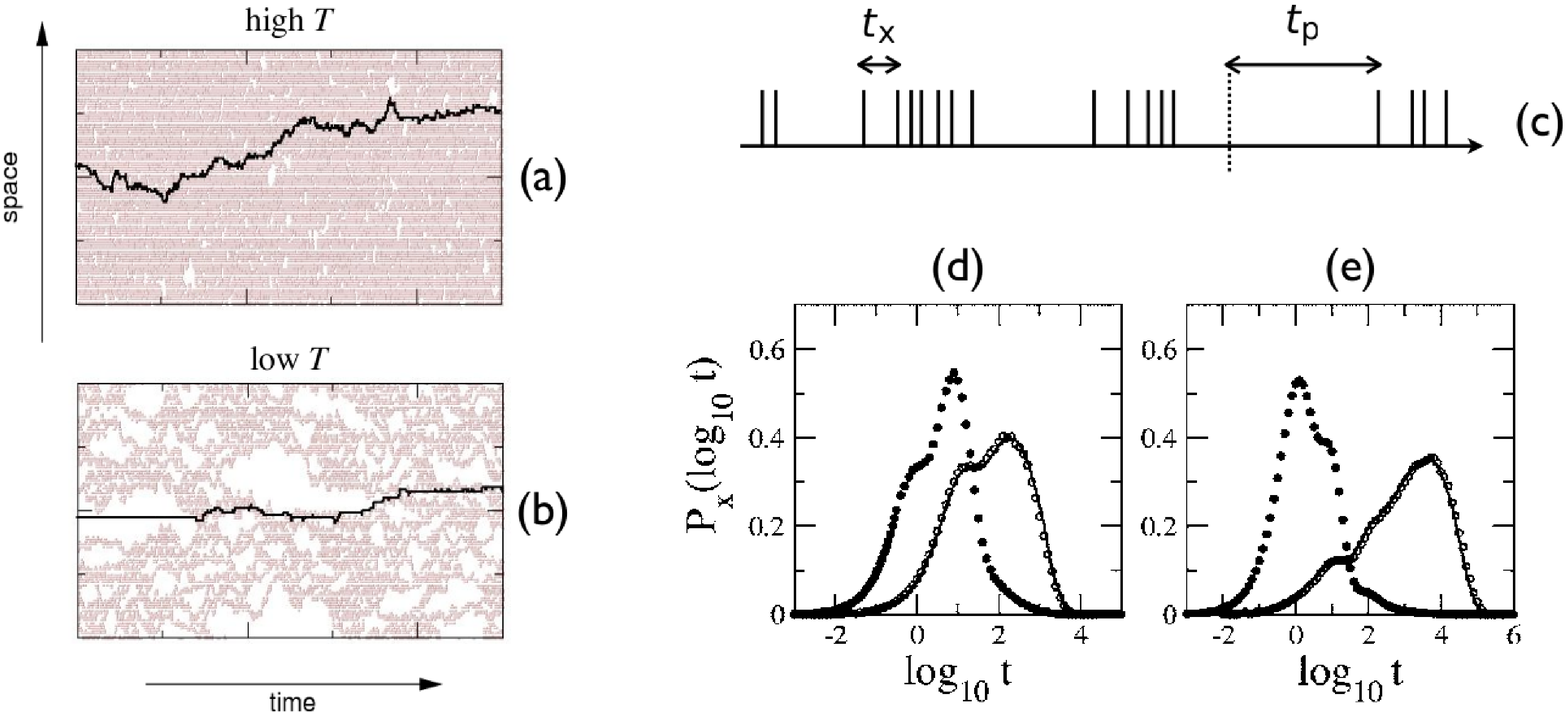}
\caption{Persistence/exchange decoupling. (a,b) Trajectory of a
probe
  particle diffusing in a KCM.  We show the case of the
  one-dimensional FA-1 for ease of illustration; similar behaviour is
  observed in other KCMs.  The probe particle can make a diffusive
  step only if allowed by the local excitations in the KCM. (Sites
  with $n_i=0$ are shown as grey in the background.) At high
  temperatures/low densities ($T=3$) diffusion becomes Fickian after short
  times and distances.  Plot (b) is on the same scale as (a), showing that
at lower temperatures ($T=0.8$)
  diffusion is intermittent and non-Fickian over much longer
  timescales and lengthscales.  (c) Timeline of displacement events.
  The waiting time between events is termed local {\em exchange} time;
  it is the time measured to the next event with the knowledge of when
  the previous one took place.  The waiting time until the next event
  from an arbitrarily chosen starting observation time is termed {\em
    persistence} time.  When the time series of events in
  non-Poissonian, as a consequence of dynamical correlation in the
  KCMs, typical persistence and exchange times are different:
  exchange times are dominated by the clustering of events, while
  persistence times are determined by the long quiescent periods.
  (d,e) The decoupling between the distributions of exchange (earlier curves) and
  persistence (later curves) times becomes more pronounced the lower the
  temperature. Results are shown for the East model, with $T=1$ in
(d) and $T=0.5$ in (e).
 Adapted from Refs.~\protect\shortcite{Jung:2004oj,Jung:2005zm}, with permission. }
\label{YJ}
\end{figure}

Four-point and similar dynamical susceptibilities measure the
dynamical fluctuations of global observables.  These observables
probe relaxation over a certain lengthscale, so it is important to
note that the scaling properties of the corresponding
susceptibilities depend on such lengthscales
\shortcite{Chandler:2006gd}. In the examples above the observable
was the persistence function, which probes relaxation on a
lengthscale of one lattice spacing. Let us now consider instead the
following susceptibility for the case of conservative (conserved
density) KCMs, $\chi_4(k,t) \equiv \lim_{k' \to 0} N^{-1} \sum_{ij}
\langle \delta{F}_i(\bm{k},t) \delta {F}_j(-\bm{k},t) \rangle
e^{i \vec{k}' \cdot (\vec{r}_i - \vec{r}_j)}$%
, where $i,j$ label the $N$ particles in the system, the position of
the $i$th particle at time $t$ is ${\bm{r}}_i(t)$, and
$\delta{F}_i(\bm{k},t)\equiv e^{i\bm{k}\cdot[{\bm{r}}_i(t)-
{\bm{r}}_i(0)]}- \langle
e^{i\bm{k}\cdot[{\bm{r}}_i(t)-{\bm{r}}_i(0)]} \rangle$
The wave vector $\bm{k}'$ appearing here  is the analogue of
$\bm{k}$ in $S_4(\bm{k},t)$ above. A normalization factor ${\mathcal
N}(t)$ could be included as before but this turns out to make no
qualitative difference. This $\chi_4$ measures the system to system
fluctuations of the self-intermediate scattering function of wave
vector $F(\bm{k},t)$.
It thus probes structural relaxation at lengthscales comparable to
$2 \pi/k$.
Figure \ref{chi4}(d) shows how $\chi_4(k,t)$ changes in behavior as
we go from large to small $k$ in the (2)-TLG: as we probe larger
lengthscales (smaller $k$) $\chi_4$ peaks at later times, and the
initial power law growth changes exponent
\shortcite{Chandler:2006gd}. This change in behavior is related to
the non-Fickian to Fickian crossover of particle diffusion
\shortcite{BerChaGar05}, which we discuss in the next subsection.

\subsection{Transport decoupling}
\label{decoupling-section}

A central consequence of DH is \index{transport decoupling}transport
decoupling.  A prominent example is \index{Stokes-Einstein
breakdown}Stokes-Einstein breakdown \shortcite{SwallenBME03}: in
deeply supercooled liquids the rate for self-diffusion is orders of
magnitude larger than what would be predicted from the
Stokes-Einstein relation between self-diffusion constant and
viscosity, $D_s \propto \eta^{-1}$.  Similar transport relations of
liquid state theory also break down near the glass transition
\shortcite{ChangS97,Ediger00}.  This is a consequence of the
dynamical fluctuations associated with DH.

A major success of KCMs is the ability to rationalize this
phenomenon. It does so in terms of the decoupling between the
different fundamental timescales for local relaxation
\shortcite{Jung:2004oj,Jung:2005zm}.  How this comes about is
illustrated in Fig. \ref{YJ}.  Panels (a) and (b) show trajectories
of a probe particle embedded in a KCM.  This can be thought of as a
molecule in a liquid that has been labelled for tracking while
coarse-graininig over the rest of the system, thus describing its
effective dynamics via a KCM.  The motion of the probe particle is
determined by the underlying fluctuations of the host KCM to which
it is coupled.  A natural dynamical rule is that the probe can make
a diffusive jump from site $i$ to site $j$ only if both sites $i$
and $j$ are excited, $n_i=n_j=0$ \shortcite{Jung:2004oj}.  Figs.\
\ref{YJ}(a) and (b) show the difference in the probe motion between
high and low temperatures.  At high $T$, Fig.\ \ref{YJ}(a),
excitations in the KCM are plentiful and probe motion appears
Brownian.  At low $T$, Fig.\ \ref{YJ}(b), excitations are scarce,
the dynamics of the KCM is heterogeneous, and probe motion is
intermittent: the probe is immobile if immersed in an inactive
space-time ``bubble''; in order to move it has to wait for an
excitation to come along.

There are two fundamental timescales that control this intermittent
motion \shortcite{Jung:2004oj,Jung:2005zm}.  The first one is the
``persistence time'', $t_{\rm p}$, that is, the time the probe needs
to wait to start moving {\em for the first time}, given an arbitrary
start time for observation.  The second timescale is the (local)
``exchange time'', $t_{\rm x}$, the time {\em between moves}.  Due
to DH jump events are not Poissonian, but display ``bunching'', and
typical persistence times can become much larger than typical
exchange times, see Fig.\ \ref{YJ}(c).  This decoupling between
persistence and exchange becomes more pronounced as temperatures is
decreased, as illustrated in Figs.\ \ref{YJ}(d) and (e).

We can approximately quantify the motion of the probe particle
\shortcite{BerChaGar05} by means of a continuous-time random walk
(CTRW) approach \shortcite{MONTROLL:1965mi}.  The probe makes random
walk steps of unit size at random times determined by the
fluctuations of the host KCM.  Lets assume that this random clock
ticks at times drawn from the exchange time distribution,
$\phi(t_{\rm x})$, but which other than that are independently
distributed.  The probability for the probe to be at position $r$ at
time $t$, or van Hove function, is $G_{s}(r,t)=\sum_{m=0}^{\infty
}\pi_{m}(t) \Gamma^{(m)}(r)$ (we assume $d=1$ for simplicity,
extension to higher dimensions is straightforward).  Here $\pi
_{m}(t)$ is the probability that the probe made $m$ steps after time
$t$, and $\Gamma^{(m)}(r)$ is the probability that a random walker
is at a distance $r$ after $m$ steps. After Laplace transforming in
time and Fourier transforming in space we obtain,
\begin{equation}
\hat{F}_s(k,\sigma) = \hat{P}(\sigma) + \cos{(k)}
\frac{\hat{p}(\sigma)}{\sigma} \frac{ 1 - \hat{\phi}(\sigma)}{ 1 -
\cos{(k)} \hat{\phi}(\sigma)} \; ,  \label{MW}
\end{equation}
where $\hat{F}_s(k,\sigma)$ is the Laplace transform of the
self-intermediate scattering function $F_s(k,t)$, which in turn is
the Fourier transform of $G_s(r,t)$. Equation (\ref{MW}) is the
Montroll-Weiss equation for the motion of the probe in the CTRW
approximation.  $\hat{\phi}(\sigma)$, $\hat{p}(\sigma)$ and
$\hat{P}(\sigma)$ are the Laplace transforms of the exchange time
distribution, the persistence time distribution, and the persistence
function $P(t)$, respectively.  The last two functions appear in
Eq.\ (\ref{MW}) because the first step is determined by the
persistence time: its distribution $p(t_{\rm p})$ is related to the
persistence function $P(t)=\pi_0(t)$ by $P(t) = \int_{t}^\infty
p(t_{\rm p}) dt_{\rm p}$. If the dynamics is stationary, then there
is nothing special about time zero and $p(t_{\rm p})= \langle
t_x\rangle^{-1}\int_{t_{\rm p}}^\infty \phi(t') dt'$, corresponding
to a uniform average over all earlier jump times; the normalization
factor is the inverse of the average exchange time $\langle
t_x\rangle = \int_0^\infty dt \phi(t) t$. From (\ref{MW}) we can
define the wavelength dependent relaxation time: $\tau(k,T) =
\lim_{\sigma \to 0}\hat{F}_s(k,\sigma)$.  We obtain
\shortcite{BerChaGar05}
\begin{equation}
\tau(k) = \tau_{\rm p} + \frac{\cos{(k)}}{1-\cos{(k)}} \tau_{\rm x}
, \label{tauk}
\end{equation}
where $\tau_{\rm x}$ and $\tau_{\rm p}$ are the average exchange and
persistence times, respectively.  At low temperatures we have
persistence/exchange decoupling, $\tau_{\rm p} \gg \tau_{\rm x}$
\shortcite{Jung:2005zm}.  The structural or alpha relaxation time is
often defined as $\tau_\alpha=\tau(k=\pi/a)$, where $a$ is the
lattice spacing which we set to unity. For these large wave vectors
the first term of (\ref{tauk}) dominates, and the structural
relaxation time is set by the persistence time, $\tau_\alpha \approx
\tau_{\rm
  p}$.  For small enough wave vectors the second term in (\ref{tauk})
dominates, $\tau(k) \approx \tau_{\rm x}/k^2$, and since the limit
of $k \to 0$ defines the diffusion rate we find that $D \approx
\tau_{\rm
  x}^{-1}$.  This explains Stokes-Einstein breakdown at low
temperatures: $D \tau_\alpha \approx \tau_{\rm p} / \tau_{\rm x}
\neq {\rm const}$ \shortcite{Jung:2004oj}.

\begin{figure}[htp]
\includegraphics[width=\columnwidth]{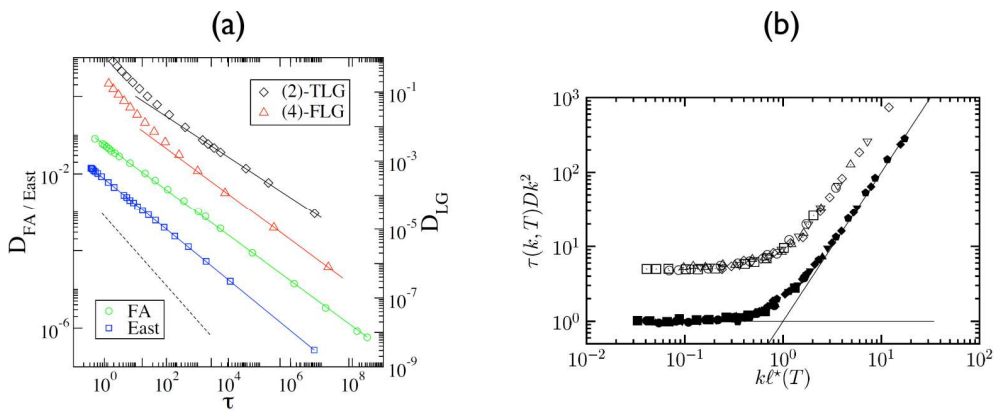}
\caption{Transport decoupling in KCMs. (a) Scaling of probe or
  self-diffusion constant with structural relaxation time
  in the one-dimensional FA-1 and East models (left
  scale) and two- and three- dimensional lattice gases (right scale).
  Solid lines indicate power law fits $D \sim
  \tau^{-\delta}$; the dashed line is the Stokes-Einstein relation $D
  \sim \tau^{-1}$.
  In all cases $\delta < 1$ at low enough $q$, indicating a breakdown
  of the Stokes-Einstein relation, as a consequence of dynamical
  fluctuations. (For the FA-1 model $\delta=1$ for dimensions larger
  than its critical dimension $d_c=2$~\protect\shortcite{Jack-Mayer-Sollich}.)
  Adapted from Refs.~\protect\shortcite{Jung:2004oj,Pan:2005lr,ashton-thesis}, with permission.
  (b) Lengthscale dependent relaxation time $\tau(k)$ in the FA-1 and
  East models.  At short lengthscales the relaxation time is $k$
  independent and determined by the persistence time, $\tau(k) \approx
  \tau_{\rm p}$.  At larger lengthscales it becomes diffusive,
  $\tau(k) \approx \tau_{\rm x}/k^2$.  This crossover is controlled by
  the Fickian lengthscale $l^* \sim \sqrt{\tau_{\rm p}/\tau_{\rm x}}$.
  The figure shows that $\tau(k)$ at different $q$ collapse under $k
  \to k l^*$.  From Ref.~\protect\shortcite{BerChaGar05}, with permission.
} \label{decoupling}
\end{figure}

The decoupling between persistence and exchange times is an effect
of dynamical fluctuations.  At low temperatures the self-diffusion
constant seems to scale as a fractional power of the relaxation time
$D \sim \tau_{\rm p}^{-\delta}$, with $\delta < 1$ (the
Stokes-Einstein relation is $\delta=1$).  This is the case for all
KCMs with strong enough constraints, as shown in Fig.\
\ref{decoupling}(a) for the FA-1 in $d=1$, the East model and
various constrained lattice gases
\shortcite{Jung:2004oj,Pan:2005lr,ashton-thesis}. The fractional
exponent $\delta \approx 0.6-0.8$ is not distinct from that observed
in experiments \shortcite{SwallenBME03,Swallen:2009ys}.

Equation (\ref{tauk}) indicates that there is a spectrum of
timescales that interpolate between a $k$ independent value
$\tau_{\rm p}$ at shorter lengthscales, to a diffusive timescale
$\tau_{\rm x}/k^2$ at large lengthscales \shortcite{BerChaGar05}.
This crossover is shown in Fig.\ \ref{decoupling}(b) for the FA-1
and East models.  It is the crossover from non-Fickian diffusion at
short lengthscales to eventual Fickian diffusion at long enough
ones.  The length $l^*$ at which this crossover takes place, or
``Fickian lengthscale", is given by $l^* \propto \sqrt{\tau_{\rm
p}/\tau_{\rm x}}$, and grows with decreasing temperature/increasing
density.  It is the distance a particle has to move before it
forgets how long it took to make the first step.  Fig.\
\ref{decoupling}(b) shows how $\tau(k)$ at different temperatures
collapse under $k \to l^* k$.  The non-Fickian to Fickian crossover
is also responsible for the wavelength dependence of four-point
functions\shortcite{Chandler:2006gd}, Fig.\ \ref{chi4}(d).

The CTRW analysis can be extended to describe the effect of driving,
for example by externally forcing the probes.  The competition
between timescales in this case leads to interesting non-linear
response behaviour, such as non-monotonic differential mobility and
giant diffusivity \shortcite{jack:011506}.  Furthermore, a study of
(\ref{MW}) in the crossover regime between non-Fickian and Fickian
explains \shortcite{Chaudhuri:2007zr} the exponential tails observed
\shortcite{Stariolo:2006ly} in van Hove functions at intermediate
times. The waiting time distributions used in the CTRW analysis
above are the ones that are obtained from the study of KCMs.  The
CTRW approach can also be used by assuming a different origin for
the waiting time distributions, as for example in the analysis of
metabasin transitions \shortcite{Heuer:2008fk} in atomistic liquids.

\begin{figure}[htp]
\includegraphics[width=\columnwidth]{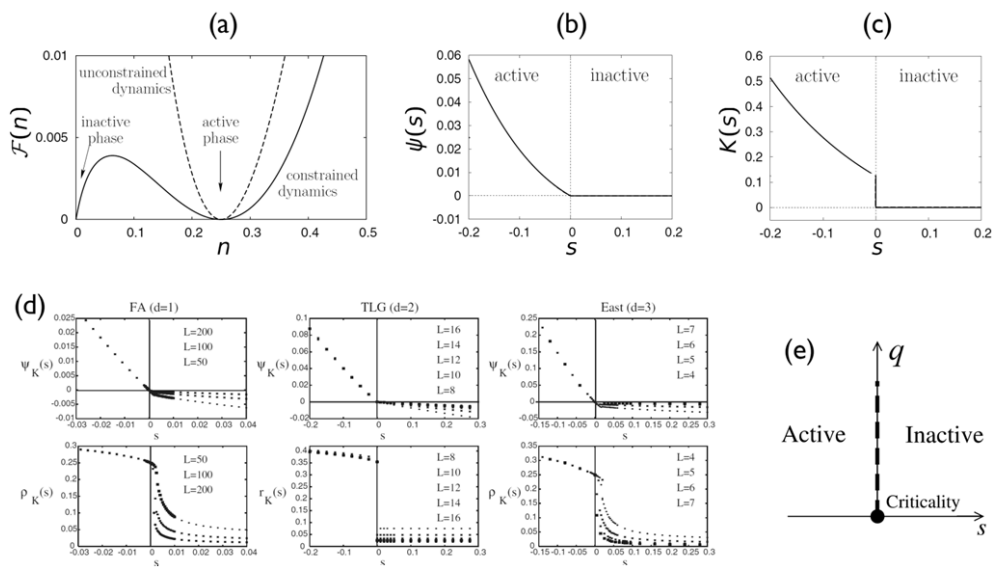}
\caption{Space-time phase transitions in KCMs.  (a) Variational
  dynamical free-energy ${\mathcal F}(n)$ in terms of mean excitation
  density. Kinetic constraints give rise to a bistable ${\mathcal F}(n)$ (full line).
  In the absence of kinetic constraints the corresponding variational
  function is unistable (dashed line).  (b) Mean-field estimate of large-deviation
  function $\psi(s)$.  There is a singularity at $s=0$, indicating a
  dynamical phase-transition.  (c) Mean activity $K(s) \equiv
  -\psi'(s)$ as a function of $s$.  This dynamical order parameter
  shows a discontinuous jump at $s=0$: the transition is a first-order
  one between an active dynamical phase and an inactive dynamical
  phase.  (d) The dynamical first-order scenario is also present in
  finite dimensions, as shown from numerical computation of $\psi(s)$
  and $K(s)$ for various KCMs.  (e) Dynamical phase diagram.  The line
  $s=0$ is one of first-order coexistence between the active dynamical
  phase ($s<0$) and the inactive one ($s>0$).  It extends all the way
  along the $q$ axis.  The critical point at $q=0$ controls the
  scaling behaviour discussed in Subsection \ref{chi4}. Adapted from
  Refs.~\protect\shortcite{Garrahan:2007la,Garrahan-et-al-2009}, with permission.}
\label{spacetime}
\end{figure}

\subsection{\index{Space-time phase transitions}Space-time phase transitions}
\label{spacetime-section}

Thermodynamically KCMs are trivial, so all interesting behaviour is
dynamical.  Nevertheless, DH pictures such as those of Fig.\
\ref{NEF-TLG} are suggestive of phase separation between two
distinct phases.  The phases of Fig.\ \ref{NEF-TLG} are
distinguished by their dynamics: dark regions are dynamically {\em
active} while light ones are dynamically {\em inactive}.
Furthermore, the phase separation in the spatial projection of low
temperature/high density equilibrium trajectories, such as those of
Fig.\ \ref{NEF-TLG}, is only mesoscopic: when
coarse-grained over large enough lengthscales, irrespective of time,
the space projected trajectories are homogeneous.  Here we show how
these observations are directly related to a true non-equilibrium
phase transition \shortcite{Garrahan:2007la}, which in contrast to
thermodynamic transitions, occurs in ensembles of trajectories and
is driven by non-equilibrium driving fields.  This phase transitions
can be studied by recourse to the large-deviation method
\shortcite{Lecomte:2007fk,Touchette:2009fk}.

A convenient order parameter to discern active and inactive dynamics
is the ``dynamical activity'' $K$ \shortcite{Garrahan:2007la},
defined as the total number of configuration changes in a
trajectory.  In a non-conservative KCM it would amount to the total
number of local changes from empty to occupied or vice versa, and in
a lattice gas (conservative KCM) to the total number of particle
displacements. The activity is extensive in space-time volume, i.e.,
typically $K = O(Nt)$, where $N$ is the number of lattice sites and
$t$ the time extension of the trajectory.  Each trajectory $x(t)$ in
the ensemble of (equilibrium) trajectories of length $t$ has a total
activity, $\hat{K}[x(t)]$.  The activity is thus distributed,
$P_t(K) = \left\langle \delta\left(K - \hat{K}[x(t)] \right)
\right\rangle$, where the average is over the set of equilibrium
trajectories, $\{ x(t) \}$.  At long enough times this probability
acquires a large-deviation form
\shortcite{Lecomte:2007fk,Touchette:2009fk}. $P_t(K) \approx e^{-t
\varphi(K/t)}$.  The function $\varphi(k)$ is called a
large-deviation function, and it plays in this dynamical context the
same role as, for example, the entropy density in the micro
canonical ensemble of equilibrium statistical mechanics.
Alternatively we can consider the generating function of $K$,
\begin{equation}
Z_t(s) \equiv \sum_K e^{-s K} P_t(K) \approx e^{t \psi(s)},
\label{Zs}
\end{equation}
which also displays a large-deviation form.  The large-deviation
function $\psi(s)$ is akin to a free-energy density, and is related
to $\varphi(K/t)$ by a Legendre transform, $\psi(s) = - \min_k
\left[
  \varphi(k)+sk \right]$.  Just like a free-energy in a thermodynamic
problem, the function $\psi(s)$ carries the information of {\em
  dynamical} phase behaviour. Specifically, its singularities
indicate {\em dynamical} phase transitions
\shortcite{Garrahan:2007la}.

The calculation of the large-deviation function $\psi(s)$ is
simplified greatly by the following observation
\shortcite{Lebowitz:1999lr}:  if ${\mathbb W}$ is the master
operator that generates the stochastic dynamics, then $\psi(s)$ is
the largest eigenvalue of a modified operator ${\mathbb W}_s$, where
${\mathbb
  W}_0={\mathbb W}$.  This reduces the calculation from that of
computing a ``partition sum'', Eq.\ (\ref{Zs}), to an eigenvalue
problem.  For example, for a spin facilitated model, ${\mathbb W}_s
= \sum_{i} f_i(\bm n) \left\{ e^{-s} \left[ (1-\rho) \sigma_i^+ +
\rho
    \sigma_i^- \right] - \rho n_i -(1-\rho) \right\}$, where $f_i(\bm
n)$ is the kinetic constraint on site $i$, and $\sigma_i^{\pm}$ are
the raising/lowering operators on site $i$.  While it is not always
possible to diagonalise such an operator analytically, bounds for
its largest eigenvalue can be estimated variationally.  For the
FA-$m$ this amounts to minimising a Landau free energy ${\mathcal F}(n)
= - n^m \left( 2 e^{-s} \sqrt{q n} - q - n \right)$.  The factor
$n^m$ comes from the kinetic constraint, and makes ${\mathcal F}(n)$
non-linear enough to allow for multiple minima, Fig.\
\ref{spacetime}(a).  The corresponding $\psi(s)$ has a singular
structure, the first derivative being discontinuous at $s=0$, Figs.\
\ref{spacetime}(b) and (c).  The meaning of this is the following.
Trajectories are organized into two dynamical phases, an active one
with $K>0$ and an inactive one with $K=0$.  The field $s$ determines
the bias for or against activity. For $s<0$ the active phase is the
dominant one, while for $s>0$ the inactive phase dominates. At $s=0$
the probability of trajectories in either phase is equal in the $t
\to \infty$ limit and we have dynamical first-order phase
coexistence \shortcite{Garrahan:2007la}.  This situation occurs in
all KCMs, Fig.\ \ref{spacetime}(d).   Interestingly, similar
dynamical phase structure is observed in spin glass models
\shortcite{PhysRevE.81.011110,PhysRevE.81.011111} and in atomistic
liquids \shortcite{Hedges:2009fk}.

Actual dynamics takes place at $s=0$.  The results above show that
this is the condition for dynamical coexistence {\em in the bulk},
i.e., infinitely far from boundaries active (ergodic) or inactive
(non-ergodic) trajectories are equally likely.  However, just like
in the case of ordinary phase transitions, boundary fields can bias
the bulk into one of the coexisting phases.  In the case of
dynamics, initial conditions play the role of a (time) boundary. In
particular, almost all possible initial configurations chosen from
the equilibrium static distribution at non-zero temperature or less
than maximal density will select the active dynamical phase.  In
this case the inactive phase manifests via rare region effects,
giving rise to DH only at the mesoscopic scale. An important open
question is whether there are physical controllable fields that play
the role of $s$ in the analysis above.

\section{Summary and outlook}
\label{conclusion}

In this chapter we have attempted to summarise recent developments
in the study of KCMs as models of glass formers.  In the long
tradition of statistical mechanics KCMs provide simplified models
that capture important ideas about the fundamental physics behind
the phenomenology of glassy systems.  Their simplicity allows for
detailed study, which in turn gives rise to further physical
insights into the glass transition problem.   The central message
from KCMs is that the complex and cooperative dynamics of glass
forming systems can be achieved without recourse to complex
thermodynamic behaviour: in KCMs thermodynamics plays essentially no
role, and complex dynamics emerge from rather simple local kinetic
rules.  These rules are local and free of disorder, but nevertheless
give rise to dynamical frustration.

The irrelevance of thermodynamics for glassy dynamics that the study
of KCMs suggest contrasts sharply with approaches such as  that of
the random first-other transition theory
\shortcite{Lubchenko:2007fk,Mezard:2000uq} where thermodynamics is
essential.  Whether thermodynamic aspects are relevant or not to
glass transition phenomena is still a matter of debate, but to the
extent that they are KCMs can say very little about them.  This can
either be seen as a flaw of the KCM based approach
\shortcite{Biroli:2005kx} or as an indication that these aspects are
described by degrees of freedom that do not contribute too much to
the long time dynamics and have therefore been coarse-grained out
\shortcite{Chandler:2005vn}. Furthermore, by their coarse-grained
and lattice based nature KCMs can in principle say very little about
short distance/short time dynamics, such as beta-relaxation or
anomalous vibrations.  There is however evidence that these short
scale phenomena are coupled to longer scale dynamic heterogeneity
(see for example \shortcite{Widmer-CooperHF04,Brito:2007uq}) so it
may be possible to capture some of these effects with
generalisations of KCMs \shortcite{Moreno:2007vn,Ashton:2009ys}.  In
any case, KCMs provide an explicitly real-space picture of glassy
dynamics.  Their main success has been the rationalization of
dynamic heterogeneity. While DH can be analysed with other
approaches, such as generalisations of mode-coupling theory
\shortcite{Biroli:2006kl}, the immediacy of the results and
explanations for DH related phenomena obtained from KCMs is
remarkable.

While we know a lot about KCMs we still do not have a satisfactory
understanding of how they emerge as an effective description from
realistic systems. (This is also the case in alternative approaches,
be it the random first-order transition
\shortcite{Lubchenko:2007fk}, or frustrated limited domains
\shortcite{Kivelson:2008lr}, where the idealised models that display
the proposed behaviour cannot be readily obtained from realistic
liquid systems.) It is usually argued \shortcite{GarCha03} that KCMs
ought to emerge from some form of local coarse-graining of a
microscopic system, but this procedure has not been shown to work
just yet (except in highly simplified situations
\shortcite{Garrahan:2000sy,Garrahan:2009fk}).   Proving a direct
connection between atomistic liquids and KCMs is arguably the
central open problem in this field.

\bibliographystyle{OUPnamed_notitle}
\bibliography{GST}

\thebibliography{0}

\bibitem[\protect\citeauthoryear{Aizenman and Lebowitz}{Aizenman and
  Lebowitz}{1988}]{AL}
Aizenman, M and Lebowitz, J~L (1988).
\newblock {\em J.\ Phys.\ A\/},~{\bf 21}, 3801--3813.

\bibitem[\protect\citeauthoryear{Aldous and Diaconis}{Aldous and
  Diaconis}{2002}]{AD}
Aldous, D and Diaconis, P (2002).
\newblock {\em J. Stat. Phys.\/},~{\bf 107}, 945--975.

\bibitem[\protect\citeauthoryear{Andersen}{Andersen}{2005}]{Andersen:2005fr}
Andersen, H~C (2005).
\newblock {\em Proc.\ Nat.\ Acad.\ Sci.\/},~{\bf 102}, 6686--6691.

\bibitem[\protect\citeauthoryear{Anderson}{Anderson}{1979}]{Anderson-LH}
Anderson, P~W (1979).
\newblock In {\em Ill-Condensed Matter} (ed. R.~Balian, R.~Maynard, and
  G.~Toulouse), Les Houches Session {XXXI}. Elsevier Science.

\bibitem[\protect\citeauthoryear{Ashton and Garrahan}{Ashton and
  Garrahan}{2009}]{Ashton:2009ys}
Ashton, D. and Garrahan, J.P. (2009).
\newblock {\em Eur.\ Phys.\ J.\ E\/},~{\bf 30}, 303--307.

\bibitem[\protect\citeauthoryear{Ashton}{Ashton}{2009}]{ashton-thesis}
Ashton, D~J (2009).
\newblock Ph.D. thesis, University of Nottingham.

\bibitem[\protect\citeauthoryear{Berthier, Chandler and Garrahan}{Berthier {\em
  et~al.}}{2005{\em a}}]{BerChaGar05}
Berthier, L, Chandler, D, and Garrahan, J~P (2005{\em a}).
\newblock {\em Europhys.\ Lett.\/},~{\bf 69}, 320--326.

\bibitem[\protect\citeauthoryear{Berthier and Garrahan}{Berthier and
  Garrahan}{2003}]{Berthier:2003lk}
Berthier, L and Garrahan, J~P (2003).
\newblock {\em Phys.\ Rev.\ E\/},~{\bf 68}, 041201.

\bibitem[\protect\citeauthoryear{Berthier and Garrahan}{Berthier and
  Garrahan}{2005}]{Berthier:2005hb}
Berthier, L and Garrahan, J~P (2005).
\newblock {\em J.\ Phys.\ Chem.\ B\/},~{\bf 109}, 3578--3585.

\bibitem[\protect\citeauthoryear{Berthier, Garrahan and Whitelam}{Berthier {\em
  et~al.}}{2005{\em b}}]{WBG}
Berthier, L, Garrahan, J~P, and Whitelam, S (2005{\em b}).
\newblock {\em Phys.\ Rev.\ E\/},~{\bf 71}, 026128--026142.

\bibitem[\protect\citeauthoryear{Bertini and Toninelli}{Bertini and
  Toninelli}{2004}]{Bertini}
Bertini, L and Toninelli, C (2004).
\newblock {\em J.\ Stat.\ Phys.\/},~{\bf 117}, 549--580.

\bibitem[\protect\citeauthoryear{Biroli, Bouchaud, Miyazaki and
  Reichman}{Biroli {\em et~al.}}{2006}]{Biroli:2006kl}
Biroli, G, Bouchaud, J~P, Miyazaki, K, and Reichman, D~R (2006).
\newblock {\em Phys.\ Rev. Lett.\/},~{\bf 97}, 195701.

\bibitem[\protect\citeauthoryear{Biroli, Bouchaud and Tarjus}{Biroli {\em
  et~al.}}{2005}]{Biroli:2005kx}
Biroli, G, Bouchaud, J~P, and Tarjus, G (2005).
\newblock {\em J.\ Chem.\ Phys.\/},~{\bf 123}.

\bibitem[\protect\citeauthoryear{Brito and Wyart}{Brito and
  Wyart}{2007}]{Brito:2007uq}
Brito, C and Wyart, M (2007).
\newblock {\em J.\ Stat.\ Mech.\ Theor.\ Exp.\/},~{\bf L08003}.

\bibitem[\protect\citeauthoryear{Buhot and Garrahan}{Buhot and
  Garrahan}{2001}]{BuhGar01}
Buhot, A and Garrahan, J~P (2001).
\newblock {\em Phys.\ Rev.\ E\/},~{\bf 64}, 021505.

\bibitem[\protect\citeauthoryear{Cancrini, Martinelli, Roberto and
  Toninelli}{Cancrini {\em et~al.}}{2007}]{CMRTjstat}
Cancrini, N, Martinelli, F, Roberto, C, and Toninelli, C (2007).
\newblock {\em J.\ Stat.\ Mech.\ Theor.\ Exp.\/},~{\bf L03001}.

\bibitem[\protect\citeauthoryear{Cancrini, Martinelli, Roberto and
  Toninelli}{Cancrini {\em et~al.}}{2008}]{CMRT}
Cancrini, N, Martinelli, F, Roberto, C, and Toninelli, C (2008).
\newblock {\em Probab.\ Theory Related Fields\/},~{\bf 140}, 459--504.

\bibitem[\protect\citeauthoryear{Cancrini, Martinelli, Roberto and
  Toninelli}{Cancrini {\em et~al.}}{2010{\em a}}]{KACMRT}
Cancrini, N, Martinelli, F, Roberto, C, and Toninelli, C (2010{\em a}).

\bibitem[\protect\citeauthoryear{Cancrini, Martinelli, Schonmann and
  Toninelli}{Cancrini {\em et~al.}}{2010{\em b}}]{CMST}
Cancrini, N, Martinelli, F, Schonmann, R~H, and Toninelli, C (2010{\em b}).

\bibitem[\protect\citeauthoryear{Cavagna}{Cavagna}{2009}]{Cavagna:2009rt}
Cavagna, A (2009).
\newblock {\em Phys.\ Rep.\/},~{\bf 476}, 51--124.

\bibitem[\protect\citeauthoryear{Chandler and Garrahan}{Chandler and
  Garrahan}{2005}]{Chandler:2005vn}
Chandler, D and Garrahan, J~P (2005).
\newblock {\em J.\ Chem.\ Phys\/},~{\bf 123}, 044511.

\bibitem[\protect\citeauthoryear{Chandler and Garrahan}{Chandler and
  Garrahan}{2010}]{GC-ARPC}
Chandler, D and Garrahan, J~P (2010).
\newblock {\em Ann.\ Rev.\ Phys.\ Chem.\/},~{\bf 61}, 191.

\bibitem[\protect\citeauthoryear{Chandler, Garrahan, Jack, Maibaum and
  Pan}{Chandler {\em et~al.}}{2006}]{Chandler:2006gd}
Chandler, D, Garrahan, J~P, Jack, R~L, Maibaum, L, and Pan, A~C (2006).
\newblock {\em Phys.\ Rev.\ E\/},~{\bf 74}, 051501.

\bibitem[\protect\citeauthoryear{Chang and Sillescu}{Chang and
  Sillescu}{1997}]{ChangS97}
Chang, I and Sillescu, H (1997).
\newblock {\em J.\ Phys.\ Chem.\ B\/},~{\bf 101}, 8794--8801.

\bibitem[\protect\citeauthoryear{Chaudhuri, Berthier and Kob}{Chaudhuri {\em
  et~al.}}{2007}]{Chaudhuri:2007zr}
Chaudhuri, P, Berthier, L, and Kob, W (2007).
\newblock {\em Phys.\ Rev.\ Lett.\/},~{\bf 99}.

\bibitem[\protect\citeauthoryear{Corberi and Cugliandolo}{Corberi and
  Cugliandolo}{2009}]{CorCug09}
Corberi, F and Cugliandolo, L~F (2009).
\newblock {\em J.\ Stat.\ Mech.\ Theor.\ Exp.\/},~{\bf P09015}.

\bibitem[\protect\citeauthoryear{Crisanti and Ritort}{Crisanti and
  Ritort}{2003}]{CriRit03}
Crisanti, A and Ritort, F (2003).
\newblock {\em J.\ Phys.\ A\/},~{\bf 36}, R181--R290.

\bibitem[\protect\citeauthoryear{Debenedetti and Stillinger}{Debenedetti and
  Stillinger}{2001}]{DebenedettiS01}
Debenedetti, P~G and Stillinger, F~H (2001).
\newblock {\em Nature\/},~{\bf 410}, 259--267.

\bibitem[\protect\citeauthoryear{Ediger}{Ediger}{2000}]{Ediger00}
Ediger, M~D (2000).
\newblock {\em Ann.\ Rev.\ Phys.\ Chem.\/},~{\bf 51}, 99--128.

\bibitem[\protect\citeauthoryear{Eisinger and J\"ackle}{Eisinger and
  J\"ackle}{1991}]{JE}
Eisinger, S and J\"ackle, J (1991).
\newblock {\em Z.\ Phys.\ B\/},~{\bf 84}, 115--124.

\bibitem[\protect\citeauthoryear{Elmatad, Chandler and Garrahan}{Elmatad {\em
  et~al.}}{2009}]{Elmatad:2009lr}
Elmatad, Y~S, Chandler, D, and Garrahan, J~P (2009).
\newblock {\em J.\ Phys.\ Chem.\ B\/},~{\bf 113}, 5563--5567.

\bibitem[\protect\citeauthoryear{Evans and Sollich}{Evans and
  Sollich}{1999}]{SE}
Evans, M~R and Sollich, P (1999).
\newblock {\em Phys.\ Rev.\ Lett\/},~{\bf 83}, 3238--3241.

\bibitem[\protect\citeauthoryear{Faggionato, Martinelli, Roberto and
  Toninelli}{Faggionato {\em et~al.}}{}]{FMRT}
Faggionato, A, Martinelli, F, Roberto, C, and Toninelli, C.
\newblock In preparation.

\bibitem[\protect\citeauthoryear{Fredrickson and Andersen}{Fredrickson and
  Andersen}{1984}]{FA}
Fredrickson, G and Andersen, H (1984).
\newblock {\em Phys.\ Rev.\ Lett.\/}, 1244--1247.

\bibitem[\protect\citeauthoryear{Fredrickson}{Fredrickson}{1988}]{Fredrickson8%
8}
Fredrickson, G~H (1988).
\newblock {\em Ann.\ Rev.\ Phys.\ Chem.\/},~{\bf 39}, 149--180.

\bibitem[\protect\citeauthoryear{Fredrickson and Brawer}{Fredrickson and
  Brawer}{1986}]{FB}
Fredrickson, G~H and Brawer, S~A (1986).
\newblock {\em J.\ Chem.\ Phys\/},~{\bf 84}, 3351.

\bibitem[\protect\citeauthoryear{Garrahan and Chandler}{Garrahan and
  Chandler}{2002}]{Garrahan:2002qp}
Garrahan, J~P and Chandler, D (2002).
\newblock {\em Phys.\ Rev.\ Lett.\/},~{\bf 89}, 035704.

\bibitem[\protect\citeauthoryear{Garrahan and Chandler}{Garrahan and
  Chandler}{2003}]{GarCha03}
Garrahan, J~P and Chandler, D (2003).
\newblock {\em Proc.\ Natl.\ Acad.\ Sci.\/},~{\bf 100}, 9710--9714.

\bibitem[\protect\citeauthoryear{Garrahan, Jack, Lecomte, Pitard, van
  Duijvendijk and van Wijland}{Garrahan {\em et~al.}}{2007}]{Garrahan:2007la}
Garrahan, J~P, Jack, R~L, Lecomte, V, Pitard, E, van Duijvendijk, K, and van
  Wijland, F (2007).
\newblock {\em Phys.\ Rev.\ Lett.\/},~{\bf 98}, 195702.

\bibitem[\protect\citeauthoryear{Garrahan, Jack, Lecomte, Pitard, van
  Duijvendijk and van Wijland}{Garrahan {\em et~al.}}{2009{\em
  a}}]{Garrahan-et-al-2009}
Garrahan, J~P, Jack, R~L, Lecomte, V, Pitard, E, van Duijvendijk, K, and van
  Wijland, F (2009{\em a}).
\newblock {\em J.\ Phys.\ A\/},~{\bf 42}, 075007.

\bibitem[\protect\citeauthoryear{Garrahan and Newman}{Garrahan and
  Newman}{2000}]{Garrahan:2000sy}
Garrahan, J~P and Newman, M E~J (2000).
\newblock {\em Phys.\ Rev.\ E\/},~{\bf 62}, 7670--7678.

\bibitem[\protect\citeauthoryear{Garrahan, Stannard, Blunt and Beton}{Garrahan
  {\em et~al.}}{2009{\em b}}]{Garrahan:2009fk}
Garrahan, J~P, Stannard, A, Blunt, M~O, and Beton, P~H (2009{\em b}).
\newblock {\em Proc.\ Nat.\ Acad.\ Sci.\/},~{\bf 106}, 15209--15213.

\bibitem[\protect\citeauthoryear{Glarum}{Glarum}{1960}]{GLARUM:1960ul}
Glarum, S~H (1960).
\newblock {\em J.\ Chem.\ Phys.\/},~{\bf 33}, 1371--1375.

\bibitem[\protect\citeauthoryear{Glotzer}{Glotzer}{2000}]{Glotzer00}
Glotzer, S~C (2000).
\newblock {\em J. Non-Cryst. Solids\/},~{\bf 274}, 342--355.

\bibitem[\protect\citeauthoryear{G{\"{o}}tze and Sj{\"{o}}gren}{G{\"{o}}tze and
  Sj{\"{o}}gren}{1992}]{GoetSjoe92}
G{\"{o}}tze, W and Sj{\"{o}}gren, L (1992).
\newblock {\em Rep.\ Prog.\ Phys.\/},~{\bf 55}, 241--376.

\bibitem[\protect\citeauthoryear{Graham, Pich\'e and Grant}{Graham {\em
  et~al.}}{1997}]{GPG}
Graham, I~S, Pich\'e, L, and Grant, M (1997).
\newblock {\em Phys.\ Rev.\ E\/},~{\bf 55}, 2132--2144.

\bibitem[\protect\citeauthoryear{Hedges and Garrahan}{Hedges and
  Garrahan}{2007}]{HedGar07}
Hedges, L~O and Garrahan, J~P (2007).
\newblock {\em J.\ Phys.\ Condens.\ Matter\/},~{\bf 19}, 205124.

\bibitem[\protect\citeauthoryear{Hedges, Jack, Garrahan and Chandler}{Hedges
  {\em et~al.}}{2009}]{Hedges:2009fk}
Hedges, L~O, Jack, R~L, Garrahan, J~P, and Chandler, D (2009).
\newblock {\em Science\/},~{\bf 323}, 1309--1313.

\bibitem[\protect\citeauthoryear{Heuer}{Heuer}{2008}]{Heuer:2008fk}
Heuer, A (2008).
\newblock {\em J.\ Phys.\ Condens.\ Matter\/},~{\bf 20}, 373101.

\bibitem[\protect\citeauthoryear{Holroyd}{Holroyd}{2003}]{Ho}
Holroyd, A~E (2003).
\newblock {\em Probab.\ Theory Related Fields\/},~{\bf 125}, 195--224.

\bibitem[\protect\citeauthoryear{J~M~Schwarz and Chayes}{J~M~Schwarz and
  Chayes}{2006}]{Schwarz}
J~M~Schwarz, A J~Liu and Chayes, L~Q (2006).
\newblock {\em Europhys.\ Lett.\/},~{\bf 73}, 560.

\bibitem[\protect\citeauthoryear{Jack and Garrahan}{Jack and
  Garrahan}{2010}]{PhysRevE.81.011111}
Jack, R~L and Garrahan, J~P (2010).
\newblock {\em Phys.\ Rev.\ E\/},~{\bf 81}, 011111.

\bibitem[\protect\citeauthoryear{Jack, Garrahan and Chandler}{Jack {\em
  et~al.}}{2006{\em a}}]{Jack:2006ph}
Jack, R~L, Garrahan, J~P, and Chandler, D (2006{\em a}).
\newblock {\em J.\ Chem.\ Phys.\/},~{\bf 125}, 184509.

\bibitem[\protect\citeauthoryear{Jack, Kelsey, Garrahan and Chandler}{Jack {\em
  et~al.}}{2008}]{jack:011506}
Jack, R~L, Kelsey, D, Garrahan, J~P, and Chandler, D (2008).
\newblock {\em Phys.\ Rev.\ E\/},~{\bf 78}, 011506.

\bibitem[\protect\citeauthoryear{Jack, Mayer and Sollich}{Jack {\em
  et~al.}}{2006{\em b}}]{Jack-Mayer-Sollich}
Jack, R~L, Mayer, P, and Sollich, P (2006{\em b}).
\newblock {\em J.\ Stat.\ Mech.\ Theor.\ Exp.\/},~{\bf P03006}.

\bibitem[\protect\citeauthoryear{J{\"{a}}ckle}{J{\"{a}}ckle}{1986}]{Jaeckle86}
J{\"{a}}ckle, J (1986).
\newblock {\em Rep.\ Prog.\ Phys.\/},~{\bf 49}, 171--231.

\bibitem[\protect\citeauthoryear{J{\"{a}}ckle and Kr{\"{o}}nig}{J{\"{a}}ckle
  and Kr{\"{o}}nig}{1994}]{JaecKro94}
J{\"{a}}ckle, J and Kr{\"{o}}nig, A (1994).
\newblock {\em J.\ Phys.\ Condens.\ Matter\/},~{\bf 6}, 7633--7653.

\bibitem[\protect\citeauthoryear{Jeng and Schwarz}{Jeng and Schwarz}{}]{Jeng}
Jeng, M and Schwarz, J~M.
\newblock Preprint arXiv:0806.1552.

\bibitem[\protect\citeauthoryear{Jung, Garrahan and Chandler}{Jung {\em
  et~al.}}{2004}]{Jung:2004oj}
Jung, YJ, Garrahan, J~P, and Chandler, D (2004).
\newblock {\em Phys.\ Rev.\ E\/},~{\bf 69}, 061205.

\bibitem[\protect\citeauthoryear{Jung, Garrahan and Chandler}{Jung {\em
  et~al.}}{2005}]{Jung:2005zm}
Jung, Y, Garrahan, J~P, and Chandler, D (2005).
\newblock {\em J.\ Chem.\ Phys.\/},~{\bf 123}, 084509.

\bibitem[\protect\citeauthoryear{Kivelson and Tarjus}{Kivelson and
  Tarjus}{2008}]{Kivelson:2008lr}
Kivelson, S~A and Tarjus, G (2008).
\newblock {\em Nature Mater.\/},~{\bf 7}, 831--833.

\bibitem[\protect\citeauthoryear{Kob and Andersen}{Kob and Andersen}{1993}]{KA}
Kob, W and Andersen, H~C (1993).
\newblock {\em Phys.\ Rev.\ E\/},~{\bf 48}, 4359--4363.

\bibitem[\protect\citeauthoryear{Lacevic, Starr, Schroder and Glotzer}{Lacevic
  {\em et~al.}}{2003}]{Lacevic:2003dz}
Lacevic, N, Starr, F~W, Schroder, T~B, and Glotzer, S~C (2003).
\newblock {\em J.\ Chem.\ Phys.\/},~{\bf 119}, 7372--7387.

\bibitem[\protect\citeauthoryear{Lebowitz and Spohn}{Lebowitz and
  Spohn}{1999}]{Lebowitz:1999lr}
Lebowitz, J~L and Spohn, H (1999).
\newblock {\em J.\ Stat.\ Phys.\/},~{\bf 95}, 333--365.

\bibitem[\protect\citeauthoryear{Lecomte, Appert-Rolland and van
  Wijland}{Lecomte {\em et~al.}}{2007}]{Lecomte:2007fk}
Lecomte, V, Appert-Rolland, C, and van Wijland, F (2007).
\newblock {\em J.\ Stat.\ Phys.\/},~{\bf 127}, 51--106.

\bibitem[\protect\citeauthoryear{L\'eonard, Mayer, Sollich, Berthier and
  Garrahan}{L\'eonard {\em et~al.}}{2007}]{LeoMaySolBerGar07}
L\'eonard, S, Mayer, P, Sollich, P, Berthier, L, and Garrahan, J~P (2007).
\newblock {\em J.\ Stat.\ Mech.\ Theor.\ Exp.\/},~{\bf P07017}.

\bibitem[\protect\citeauthoryear{Lubchenko and Wolynes}{Lubchenko and
  Wolynes}{2007}]{Lubchenko:2007fk}
Lubchenko, V and Wolynes, P~G (2007).
\newblock {\em Ann.\ Rev.\ Phys.\ Chem.\/},~{\bf 58}, 235--266.

\bibitem[\protect\citeauthoryear{Mayer, Leonard, Berthier, Garrahan and
  Sollich}{Mayer {\em et~al.}}{2006}]{MayLeoBerGarSol06}
Mayer, P, Leonard, S, Berthier, L, Garrahan, J~P, and Sollich, P (2006).
\newblock {\em Phys.\ Rev.\ Lett.\/},~{\bf 96}, 030602.

\bibitem[\protect\citeauthoryear{Mayer and Sollich}{Mayer and
  Sollich}{2007}]{MaySol07}
Mayer, Peter and Sollich, Peter (2007).
\newblock {\em J.\ Phys.\ A\/},~{\bf 40}, 5823--5856.

\bibitem[\protect\citeauthoryear{Merolle, Garrahan and Chandler}{Merolle {\em
  et~al.}}{2005}]{Merolle:2005ix}
Merolle, M, Garrahan, J~P, and Chandler, D (2005).
\newblock {\em Proc.\ Nat.\ Acad.\ Sci.\/},~{\bf 102}, 10837--10840.

\bibitem[\protect\citeauthoryear{Mezard and Parisi}{Mezard and
  Parisi}{2000}]{Mezard:2000uq}
Mezard, M and Parisi, G (2000).
\newblock {\em J.\ Phys.\ Condens.\ Matter\/},~{\bf 12}, 6655--6673.

\bibitem[\protect\citeauthoryear{Montroll and Weiss}{Montroll and
  Weiss}{1965}]{MONTROLL:1965mi}
Montroll, E~W and Weiss, G~H (1965).
\newblock {\em J.\ Math.\ Phys.\/},~{\bf 6}, 167.

\bibitem[\protect\citeauthoryear{Moreno and Colmenero}{Moreno and
  Colmenero}{2007}]{Moreno:2007vn}
Moreno, A~J and Colmenero, J (2007).
\newblock {\em J.\ Phys.\ Condens.\ Matter\/},~{\bf 19}, 205144.

\bibitem[\protect\citeauthoryear{Palmer}{Palmer}{1989}]{Palmer89}
Palmer, R~G (1989).
\newblock In {\em Cooperative Dynamics in Complex Physical Systems} (ed.
  H.~Takayama), Volume~43, Springer Series in Synergetics, pp.\  118--127.
  Springer, Heidelberg.

\bibitem[\protect\citeauthoryear{Palmer, Stein, Abrahams and Anderson}{Palmer
  {\em et~al.}}{1984}]{PalmerSAA84}
Palmer, R~G, Stein, D~L, Abrahams, E, and Anderson, P~W (1984).
\newblock {\em Phys.\ Rev.\ Lett.\/},~{\bf 53}, 958--961.

\bibitem[\protect\citeauthoryear{Pan, Garrahan and Chandler}{Pan {\em
  et~al.}}{2005}]{Pan:2005lr}
Pan, A~C, Garrahan, J~P, and Chandler, D (2005).
\newblock {\em Phys.\ Rev.\ E\/},~{\bf 72}, 041106.

\bibitem[\protect\citeauthoryear{Reiter}{Reiter}{1991}]{Reiter}
Reiter, J (1991).
\newblock {\em J.\ Chem.\ Phys.\/},~{\bf 95}, 544--554.

\bibitem[\protect\citeauthoryear{Reiter, Mauch and J{\"{a}}ckle}{Reiter {\em
  et~al.}}{1992}]{ReiMauJaec92}
Reiter, J, Mauch, F, and J{\"{a}}ckle, J (1992).
\newblock {\em Physica A\/},~{\bf 184}, 458--476.

\bibitem[\protect\citeauthoryear{Ritort and Sollich}{Ritort and
  Sollich}{2003}]{RitSol03}
Ritort, F and Sollich, P (2003).
\newblock {\em Adv.\ Phys.\/},~{\bf 52}, 219--342.

\bibitem[\protect\citeauthoryear{S~Butler}{S~Butler}{1991}]{BH}
S~Butler, P~Harrowell (1991).
\newblock {\em J.\ Chem.\ Phys.\/},~{\bf 85}, 4466.

\bibitem[\protect\citeauthoryear{S~N~Dorogovtsev and Mendes}{S~N~Dorogovtsev
  and Mendes}{2006}]{Mendes}
S~N~Dorogovtsev, A V~Goltsev and Mendes, J F~F (2006).
\newblock {\em Phys.\ Rev.\ Lett.\/},~{\bf 96}, 040601.

\bibitem[\protect\citeauthoryear{Schonmann}{Schonmann}{1992}]{Schonman}
Schonmann, R~H (1992).
\newblock {\em Ann.\ Probab.\/},~{\bf 20}, 174--193.

\bibitem[\protect\citeauthoryear{Sellitto, Biroli and Toninelli}{Sellitto {\em
  et~al.}}{2005}]{Sellitto}
Sellitto, M, Biroli, G, and Toninelli, C (2005).
\newblock {\em Europhys.\ Lett.\/},~{\bf 69}, 496.

\bibitem[\protect\citeauthoryear{Sollich and Evans}{Sollich and
  Evans}{1999}]{Sollich:1999tg}
Sollich, P and Evans, M~R (1999).
\newblock {\em Phys.\ Rev.\ Lett.\/},~{\bf 83}, 3238--3241.

\bibitem[\protect\citeauthoryear{Sollich and Evans}{Sollich and
  Evans}{2003}]{SolEva03}
Sollich, P and Evans, M~R (2003).
\newblock {\em Phys.\ Rev.\ E\/},~{\bf 68}, 031504.

\bibitem[\protect\citeauthoryear{Stariolo and Fabricius}{Stariolo and
  Fabricius}{2006}]{Stariolo:2006ly}
Stariolo, D~A and Fabricius, G (2006).
\newblock {\em J.\ Chem.\ Phys.\/},~{\bf 125}.

\bibitem[\protect\citeauthoryear{Swallen, Bonvallet, McMahon and
  Ediger}{Swallen {\em et~al.}}{2003}]{SwallenBME03}
Swallen, S~F, Bonvallet, P~A, McMahon, R~J, and Ediger, M~D (2003).
\newblock {\em Phys.\ Rev.\ Lett.\/},~{\bf 90}, 015901.

\bibitem[\protect\citeauthoryear{Swallen, Traynor, McMahon, Ediger and
  Mates}{Swallen {\em et~al.}}{2009}]{Swallen:2009ys}
Swallen, S~F, Traynor, K, McMahon, R~J, Ediger, M~D, and Mates, T~E (2009).
\newblock {\em J.\ Phys.\ Chem.\ B\/},~{\bf 113}, 4600--4608.

\bibitem[\protect\citeauthoryear{Toninelli and Biroli}{Toninelli and
  Biroli}{2004}]{TB}
Toninelli, C and Biroli, G (2004).
\newblock {\em J.\ Stat.\ Phys.\/},~{\bf 117}, 27--54.

\bibitem[\protect\citeauthoryear{Toninelli and Biroli}{Toninelli and
  Biroli}{2008{\em a}}]{TBspiral1}
Toninelli, C and Biroli, G (2008{\em a}).
\newblock {\em J.\ Stat.\ Phys.\/},~{\bf 130}, 83--112.

\bibitem[\protect\citeauthoryear{Toninelli and Biroli}{Toninelli and
  Biroli}{2008{\em b}}]{TBspiral2}
Toninelli, C and Biroli, G (2008{\em b}).
\newblock {\em Eur.\ Phys.\ J.\ B\/},~{\bf 130}.

\bibitem[\protect\citeauthoryear{Toninelli, Biroli and Fisher}{Toninelli {\em
  et~al.}}{2004{\em a}}]{KAletter}
Toninelli, C, Biroli, G, and Fisher, D~S (2004{\em a}).
\newblock {\em Phys.\ Rev.\ Lett.\/},~{\bf 92}, 185504.

\bibitem[\protect\citeauthoryear{Toninelli, Biroli and Fisher}{Toninelli {\em
  et~al.}}{2004{\em b}}]{ToninelliBF04}
Toninelli, C, Biroli, G, and Fisher, D~S (2004{\em b}).
\newblock {\em Phys.\ Rev.\ Lett.\/},~{\bf 92}.

\bibitem[\protect\citeauthoryear{Toninelli, Biroli and Fisher}{Toninelli {\em
  et~al.}}{2005{\em a}}]{KAnoi}
Toninelli, C, Biroli, G, and Fisher, D~S (2005{\em a}).
\newblock {\em J.\ Stat.\ Phys.\/},~{\bf 120}, 167--238.

\bibitem[\protect\citeauthoryear{Toninelli, Biroli and Fisher}{Toninelli {\em
  et~al.}}{2006}]{knight}
Toninelli, C, Biroli, G, and Fisher, D~S (2006).
\newblock {\em Phys.\ Rev.\ Lett.\/},~{\bf 96}, 035702.

\bibitem[\protect\citeauthoryear{Toninelli, Biroli and Fisher}{Toninelli {\em
  et~al.}}{2007}]{TBFreply}
Toninelli, C, Biroli, G, and Fisher, D~S (2007).
\newblock {\em Phys.\ Rev.\ Lett.\/},~{\bf 98}, 129602.

\bibitem[\protect\citeauthoryear{Toninelli, Wyart, Berthier, Biroli and
  Bouchaud}{Toninelli {\em et~al.}}{2005{\em b}}]{ToninelliWBBB05}
Toninelli, C, Wyart, M, Berthier, L, Biroli, G, and Bouchaud, J~P (2005{\em
  b}).
\newblock {\em Phys.\ Rev.\ E\/},~{\bf 71}, 041505.

\bibitem[\protect\citeauthoryear{Touchette}{Touchette}{2009}]{Touchette:2009fk}
Touchette, H (2009).
\newblock {\em Phys.\ Rep.\/},~{\bf 478}, 1--69.

\bibitem[\protect\citeauthoryear{van Duijvendijk, Jack and van Wijland}{van
  Duijvendijk {\em et~al.}}{2010}]{PhysRevE.81.011110}
van Duijvendijk, K, Jack, R~L, and van Wijland, F (2010).
\newblock {\em Phys.\ Rev.\ E\/},~{\bf 81}, 011110.

\bibitem[\protect\citeauthoryear{Whitelam, Berthier and Garrahan}{Whitelam {\em
  et~al.}}{2004}]{Whitelam:2004wo}
Whitelam, S, Berthier, L, and Garrahan, J~P (2004).
\newblock {\em Phys.\ Rev.\ Lett.\/},~{\bf 92}, 185705.

\bibitem[\protect\citeauthoryear{Whitelam and Garrahan}{Whitelam and
  Garrahan}{2004}]{Whitelam:2004wt}
Whitelam, S and Garrahan, J~P (2004).
\newblock {\em J.\ Phys.\ Chem.\ B\/},~{\bf 108}, 6611--6615.

\bibitem[\protect\citeauthoryear{Widmer-Cooper, Harrowell and
  Fynewever}{Widmer-Cooper {\em et~al.}}{2004}]{Widmer-CooperHF04}
Widmer-Cooper, A., Harrowell, P., and Fynewever, H. (2004).
\newblock {\em Phys. Rev. Lett.\/},~{\bf 93}, 135701.

\endthebibliography
\end{document}